\documentclass{aa}  

\usepackage{natbib}
\bibpunct{(}{)}{;}{a}{}{,} 
\usepackage{color}
\usepackage{graphicx}
\usepackage{txfonts}
\usepackage[multiple]{footmisc}
\begin{document}

   \title{Coronal magnetic field evolution over the cycle 24}
   \author{I. Chifu
          \inst{1,2}
          \and
          B. Inhester\inst{1}
          \and
          T. Wiegelmann\inst{1}
          }

   \institute{Max-Planck-Institut f\"ur Sonnensystemforschung, Justus-von-Liebig-Weg 3,
37077 G\"ottingen, Germany;\\
              \email{chifu@mps.mpg.de}
         \and
        Institute of Astrophysics and Geophysics, University of G\"ottingen, Friedrich-Hund-Platz 1 37077 G\"ottingen, Germany
             }


 
  \abstract
   {The photospheric magnetic field vector is continuously derived from measurements, while reconstruction of the three-dimensional (3D) coronal magnetic field requires modelling with photospheric measurements as a boundary condition. For decades the cycle variation of the magnetic field in the photosphere has been investigated. To present, there is no study to show the evolution of the coronal magnetic flux in the corona, neither the evolution of solar cycle magnetic free energy. }
   {The paper aims to analyze the temporal variation of the magnetic field and free magnetic energy in the solar corona for the solar cycle 24 and how the magnetic field behaves in the two hemispheres. We want to investigate if we can obtain better estimates of the magnetic field at Earth using the nonlinear force-free field extrapolation method.
}
   {To model the magnetic field over cycle 24 we apply the nonlinear force-free field (NLFFF) optimization method to the entire set of the synoptic vector magnetic maps derived from the observations of Heliospheric and Magnetic Imager (HMI) onboard Solar Dynamic Observatory (SDO).}
   {From our results we found that during the solar cycle 24, the maximum of the Sun's dynamics is different than the sunspot number (SSN) maximum peak. The major contribution to the total unsigned flux is provided by the flux coming from the magnetic field structures other than sunspots (MSOS) within latitudes of -30$^\circ$ and +$30^\circ$. The magnetic flux variation during the solar cycle 24 shows a different evolution in the corona than in the photosphere. We found a correlation value of 0.8 between the derived magnetic energy from our model and the flare energy index derived from observations. On average, the cycle 24 had a higher number of sunspots in the northern hemisphere (NH) but stronger flux in the southern hemisphere (SH) which could more effectively reach the higher layers of the atmosphere. The coupling between the hemispheres increases with height. The strongest asymmetries in the unsigned magnetic flux are between the two SSN peaks.}
   {}


   \maketitle


\section{Introduction}

In 1843, the German astronomer Schwabe discovered the 11-years solar cycle. A few years later, Wolf compiled Schwabe's observations of sunspots with other observations dating back to Galileo's first recordings. Based on Wolf's numbering, cycle number one falls into the period 1755-1766. The connection of the solar cycle with a variation in the Sun's magnetic activity was only recognised at the beginning of 20$^{th}$ century by the American astronomer Hale \citep[see][for a review paper]{2010LRSP....7....1H}.

Investigating the cycle variation in more detail, it was found that the SSN for many maxima displays a double peak known as Gnevyshev peak \citep{1977SoPh...51..175G,1963SvA.....7..311G,2018ApJ...866...17K}. The period between the double two peaks is known as Gnevyshev gap. It was noted that the two peak phenomena seem to appear in each of the hemispheres \citep{2010SoPh..261..193N,2006A&A...447..735T}. One possible explanation was suggested by \citet{2018ApJ...866...17K} based on dynamo simulations. The authors conclude that the double peaks of one cycle are due to surges of opposite polarity in the polar surface magnetic field observed in the previous cycle. 
The global magnetic field variation during a solar cycle can be related to the coupling between the northern and southern hemispheres. The coupling strength then determines the phase-lag between them \citep{2010SoPh..261..193N}.  

The effects of the solar cycle in the corona did not receive as much attention as those in the photosphere. Probably the main reason is the difficulty in measuring the magnetic field in the corona. While the vector magnetic field in the photosphere is derived from daily observations made by HMI/SDO, the magnetic field in the corona has occasionally been derived from direct measurements. Instead, an estimate of the coronal magnetic field can be obtained through extrapolation from the magnetic field in the photosphere.

The variation of the sunspot activity during the solar cycle certainly has an impact on the solar corona. The strong magnetic flux from larger sunspots shapes the magnetic field in most of the corona. The contribution from numerous small scale flux concentrations in quiet soar surface regions is less obvious \citep{2010ApJ...723L.164L}. While the effect of the concentrated flux in the sunspots can be seen clearly in the extreme ultra-violet (EUV) coronal images, the effect of the flux from smaller regions is not as visible in the corona. The reduced spacial scale data used as input in the NLFFF extrapolations does not allow us to see if there are any effects in the corona due to small area photospheric magnetic flux concentrations.

One of the few investigations of the solar cycle variations in the corona deals with the temperature and emission measure using EUV data from 2010-2017, see \cite{2017SciA....3E2056M}. The authors compared these thermal plasma  properties with variations in the surface magnetic field and concluded that their analysis could be improved if the comparison was made with the coronal magnetic field modelled over long times periods of, for example, a solar cycle. 

In this paper, we pick up the suggestion by \cite{2017SciA....3E2056M} and model the coronal magnetic field for the entire solar cycle 24. To our knowledge, there has been no comparable investigation of the temporal evolution of the magnetic flux and magnetic energy in the solar corona during an entire solar cycle before. We consider the above quantities in each hemisphere separately to see how the hemispheric balance changes from photosphere to the corona.

The solar cycle 24 appears to be the weakest compared with the previous four cycles which were already weak. \cite{2018A&A...618A.148J} reports an unusual polar field reversal during cycle 24. It is therefore not obvious to which extent our results can be generalised to previous solar cycles.

The paper is structured as follows: in Section 2, we briefly describe the extrapolation method used for deriving the coronal magnetic field; in Section 3, we describe the data and the set-up of the extrapolation. The results are presented in Section 4, followed by discussions and conclusions in Section 5.

\section{The coronal magnetic field extrapolation method}
A basic assumption of the extrapolation method is the stationarity of the coronal magnetic field and its photospheric boundary on scales larger than the spatial resolution of the field model and on temporal scales longer than an Alfven transit time through the model domain. Both these assumptions hold since the spatial and temporal scale we consider in this study are large, a fraction of the solar radius and a fraction of the 11-year cycle, respectively. The other major assumption is that non-magnetic forces such as thermal and dynamic pressure and gravitational forces can be neglected. The ratio of the thermal pressure to the magnetic pressure is expressed by the plasma beta parameter ($\beta$). According to a study by \cite{ Gary2001}, in the corona, the magnetic pressure dominates the plasma pressure over a large range of coronal altitudes. Gravitational forces and dynamic pressure can be neglected on the scales considered here where the gravitational escape speed and the plasma flow speed stays well below the Alfven speed. 

For a review on global coronal magnetic field extrapolations, including a discussion of the effects of neglecting terms from a full magnetohydrodynamic (MHD) model effect can be found in \cite{2017SSRv..210..249W}. The work done in our current paper is based on global NLFFF-extrapolations. A source for evaluating both the problems and achievements of NLFFF on global scales in comparison with other nonpotential global models is the study by \citet{2018SSRv..214...99Y}. In this paper, which was based on two ISSI\footnote{International Space Science Institute, Bern, Switzerland}-meetings, a number of seven different global nonpotential codes (two NLFFF-codes, magnetohydrostatics(MHS), force-free electrodynamics, magneto frictional, MHD, zero-beat MHD) have been applied to model the solar corona during the solar eclipse from March 2015, and the results have been compared. The authors found that in all investigated models some details match with observations, while others do not. A main reason for this is related to the distribution of electric currents, because currents are a key aspect regarding the nonpotentiality of the coronal magnetic field. It was concluded that the NLFFF extrapolations performed best in active regions (AR) where reliable vector magnetic fields are measured. Such data are not available outside AR and to compute the free energy here requires time-evolving models, as for example the magneto-frictional approach. As a result, the global NLFFF-approach is likely to underestimate the amount of free energy, because it neglects energy sources outside ARs. 
\cite{2015A&A...584A..68P} questioned the assumption that the plasma $\beta$ is orders of magnitudes lower than unity in the corona, a basic assumption of the NLFFF extrapolations. By performing MHD simulations, which are also an approximation, the authors concluded that the low plasma $\beta$ assumption in the corona does not hold, and plasma $\beta$ can be larger than $0.1$. They further concluded that neglecting the finite-beta effects in NLFFF makes estimations of the free magnetic energy unreliable if the plasma $\beta$ is of the order of the free energy. This effect is strongest for ARs with relatively small free energy ($5-10 \%$ of the potential field energy). For ARs with strong parallel electric currents and high relative magnetic energy the effect becomes lower. Another effect is that the spatial resolution of the computation matters and higher resolution computations show more free magnetic energy \citep[see][for details]{2015ApJ...811..107D}. As a consequence, these three effects (missing vector magnetograms in the quiet Sun, neglecting finite-beta effects, limited spatial resolution) cause an underestimation of the free magnetic energy in global NLFFF-models. The free energies deduced from these models can be considered as a lower limit of the magnetic energy. Nevertheless, despite these shortcomings, the amount of energy computed from NLFFF is a useful quantity. Simpler potential field source surface models have a free magnetic energy of zero by construction and even if NLFFF-models provide only a lower limit of the access energy, this can give some insight regarding energy sources available for the activity of the Sun. \cite{2009ApJ...696.1780D} address the limitation of NLFFF-methods when applied to an AR in a small computational box and they suggested that the extrapolation of an AR should be performed using a much larger field of view (FOV) around the target. Such data are now (since 2010) available from SDO/HMI and the global computations based on synoptic vector maps have the largest possible FOV. It was also pointed out that the NLFFF-codes should incorporate measurement errors in the vector fields, which current state-of-the-art NLFFF codes do. Another identified issue was that the traced magnetic field lines from the solution of the extrapolation do not always match perfectly the observations and that the photosphere-corona interface is not well understood. Progress regarding a better modelling of this forced interface region has been done with a MHS model, see for example \cite{2020A&A...640A.103Z}. MHS models take non-magnetic forces and finite-beta effects self-consistently into account, but for a physical understanding of the thin layer between photosphere and corona, very high resolution vector magnetograms are required and  MHS-models need a much longer computing time than NLFFF. Currently, the resolution of the available synoptic vector magnetograms is not sufficient and consequently this approach is still not feasible for global models. For more details about force-free coronal magnetic fields see the review papers by \cite{2012LRSP....9....5W, 2021LRSP...18....1W}.

The extrapolation method we use was initially proposed by \cite{WheatlandEtal2000} and implemented by \cite{Wiegelmann04}. It is described in detail in \cite{2007SoPh..240..227W} so we will only mention the special variant which we used for this study. The method is based on the force-free field assumption, that is the Lorentz force $\mathbf{j} \times \mathbf{B} = 0$ vanishes together with $\nabla \cdot \mathbf{B}=0$. This assumption is an idealization as it implies stationarity and the vanishing of pressure and gravity forces. The coronal field $\mathbf{B}$ in domain V is obtained by minimising simultaneously the three integral terms from the functional $L=L_\mathrm{1}+L_\mathrm{2}+L_\mathrm{3}$, where 

\begin{gather}
  L_\mathrm{1}=\int_V w_\mathrm{f} \frac{|(\nabla \times \Vec{B}) \times
    \Vec{B}|^2}{B^2} \;r^2 \sin \theta \;\mathrm{d}r \;\mathrm{d}\theta \;\mathrm{d}\phi,
  \label{L1}\\
  L_\mathrm{2}=\int_V w_\mathrm{d} |\nabla \cdot \Vec{B}|^2 \;r^2 \sin \theta \;\mathrm{d}r \;\mathrm{d}\theta \;\mathrm{d}\phi,
  \label{L2}\\
  L_\mathrm{3}=\int_S ( \Vec{B} - \Vec{B}_\mathrm{obs} )
           \cdot \mathrm{diag(\sigma^{-2}_\mathrm{\alpha})}
           \cdot (\mathbf{B}-\mathbf{B}_\mathrm{obs}) \;r^2 \sin \theta \;\mathrm{d}\theta \;\mathrm{d}\phi,
  \label{L3}
\end{gather}

which intend to cause the resulting field to be force-free (term $L_{\ref{L1}}$), divergence-free (term $L_{\ref{L2}}$) and to match the photospheric boundary observations $\Vec{B}_{\mathrm{obs}}$ (term $L_{\ref{L3}}$). Here, $w_{\mathrm{d}}$ is a space-dependent weight which tapers the lateral and top boundaries \citep[see][for more details]{Wiegelmann04}, $\sigma^{-2}_{\mathrm{\alpha}}$ is the space-dependent variance of the measurements error in the surface field $\mathbf{B}_{\mathrm{obs}}$ at the bottom boundary S of V \citep[see][for more details]{2014A&A...562A.105T}. 

One of the main assumptions of the force-free magnetic field modelling is that the plasma $\beta$ is much, or at least one order of magnitude, lower than unity. In the corona, this assumption is valid \cite[see][]{ Gary2001}, but in the photosphere, plasma $\beta$ is of the order of unity. According to \cite{2006SoPh..233..215W}(and the references therein), a vector magnetogram fulfils the force-free assumption if the total flux, force, and torque are much lower than unity \cite[see][for the expression of these quantities]{Tadesse2011}. If this condition is not fulfilled, one should apply preprocessing on the vector magnetograms \citep{2006SoPh..233..215W,Tadesse2011}.
\cite{2012SoPh..281...37W} checked the consistency of the HMI vector magnetogram with the force-free assumption by calculating the expressions of force, torque and flux. For the evaluation, they used a FOV that comprised an AR and its surroundings but not the entire vector magnetogram. They concluded that for the investigated data set, the HMI vector magnetograms do not need preprocessing because the values of the force, torque and flux were of the order of 10$^{-2}$, 10$^{-3}$, respectively. Nevertheless, one should always check the consistency of the boundary data.

The computational domain V extends over $r$ = [1, 2.5] solar radius ($R_\mathrm{s}$) in heliocentric radial direction, $\theta = [-70^\circ,70^\circ]$ in latitude and $\phi = [0^\circ,360^\circ]$ in longitude. At altitudes beyond 2.5 $R_\mathrm{s}$, the plasma $\beta$ increases toward unity and dynamic forces can no longer be neglected as the solar wind bulk velocity approaches the Alfv\`en speed. The latitudinal boundaries exclude the poles areas because the surface data in polar latitudes has only poor quality and the numerical finite-difference representation we use for the integrals in (\ref{L1} - \ref{L3}) becomes singular at the poles. The method uses the observed vector magnetogram as the bottom boundary. The lateral and top boundaries are fixed from an initially calculated potential field.

The NLFF field reconstructions are calculated iteratively from an initial magnetic field until the field \Vec{B} has relaxed to a force-free state. We use the multiscale approach. On the coarsest grid of 45$\times$70$\times$180 in $r\times\theta\times\phi$, we use an initial potential field determined from the normal component of the surface field. The solution of the NLFFF extrapolation, on any but the finest grid, is interpolated on the next grid and used as the initial condition for the extrapolation on the new grid. The final coronal field model analysed below is the result obtained on the 180$\times$280$\times$720 grid. At any level change the grid size is reduced (enhanced) by a factor of two.


\section{The data}

As the bottom boundary input for the NLFFF extrapolation, we used vector magnetic field data from the HMI instrument on board SDO \citep{2012SoPh..275....3P}. 
One of the data products suitable for the full Sun extrapolation provided by the HMI team \citep{Liu2017} is the synoptic maps which are derived from the daily HMI vector magnetograms. Each longitude stripe from the synoptic map represents the average of 20 magnetograms obtained at the central meridian passage of that longitude. The size of the HMI synoptic magnetograms is 3600x1440 pixels and the pixel size is 0.1 deg in longitude and 0.001 in sin latitude. 

\begin{figure}[ht]
  \centering
   \includegraphics[width=9.5cm, trim = 30 290 80 280, clip]{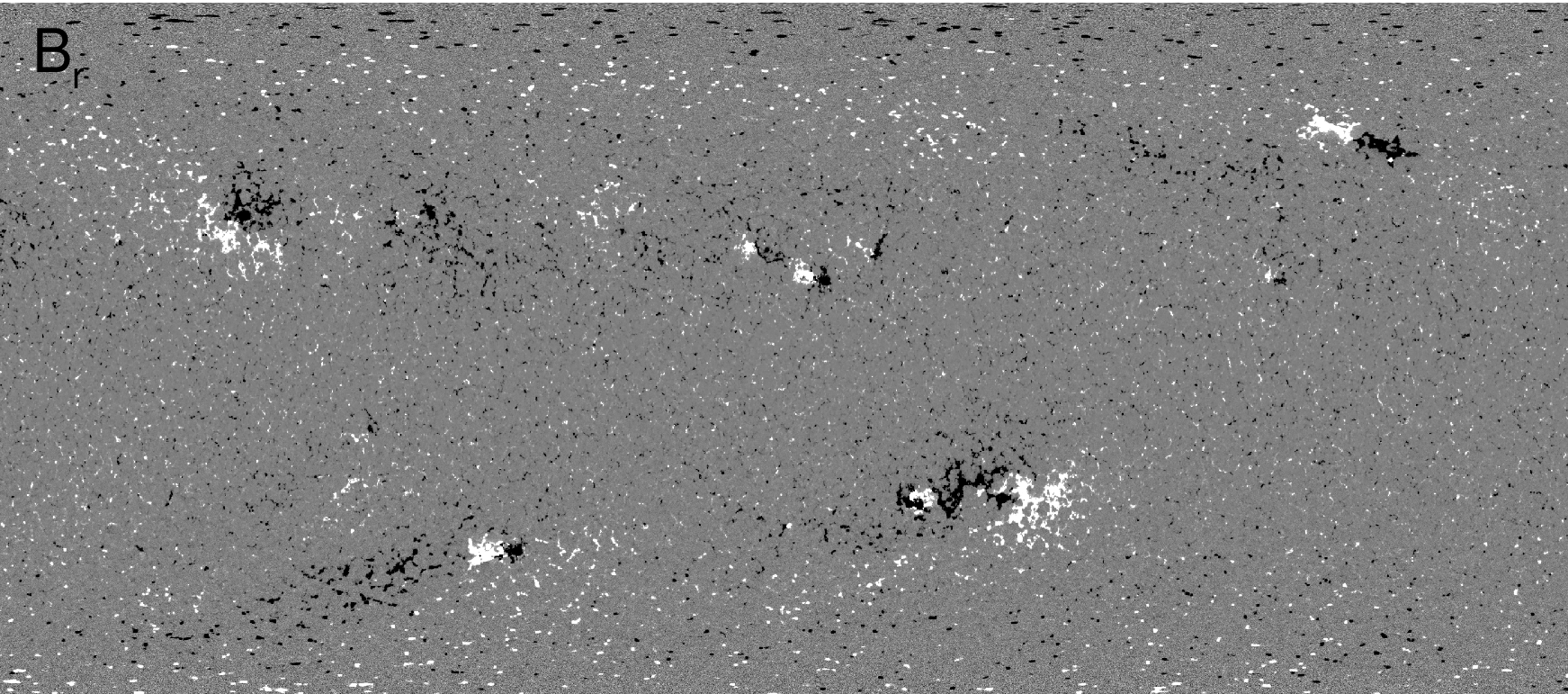} \\
 \includegraphics[width=9.5cm, trim = 30 290 80 280, clip]{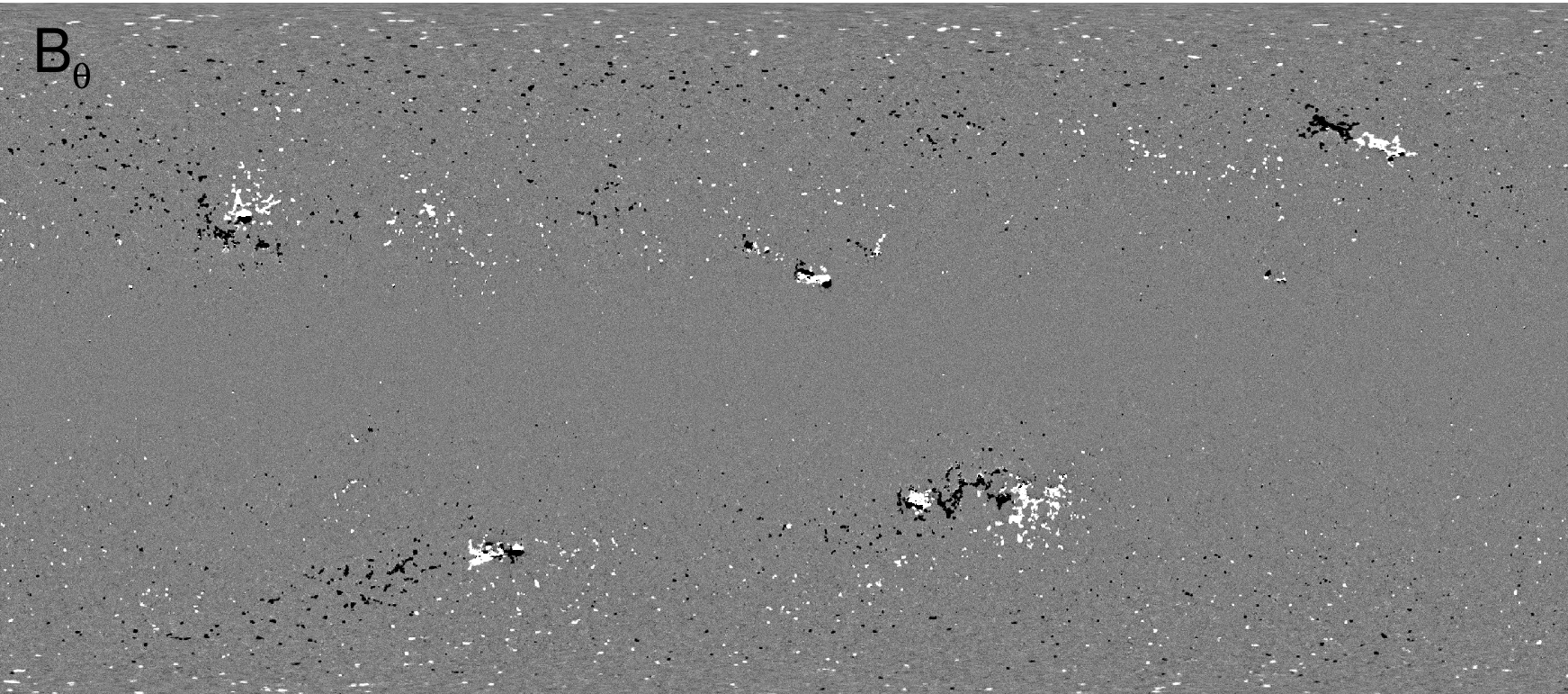} \\ 
  \includegraphics[width=9.5cm, trim = 30 290 80 280, clip]{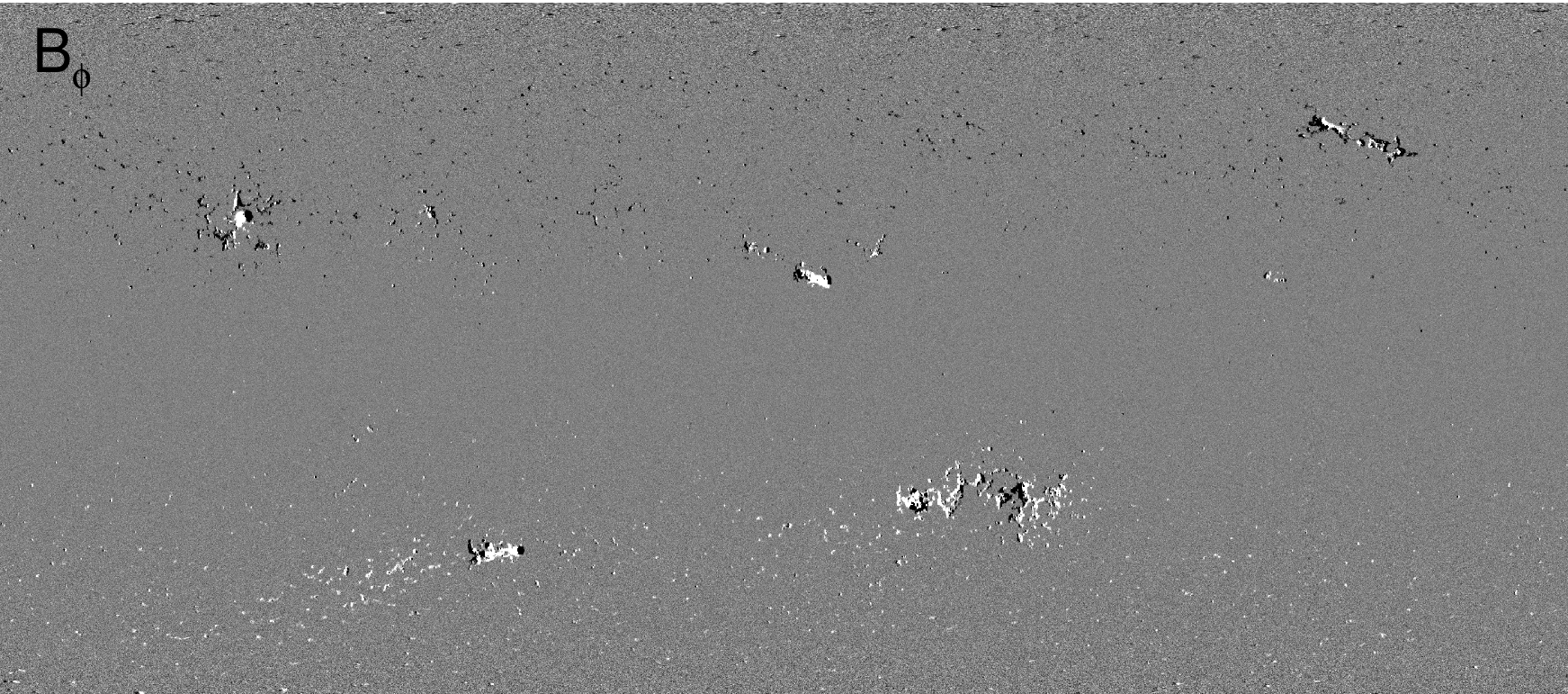}
  \caption{HMI synoptic vector magnetogram for the CR 2103 showing all of the three components of the magnetic field.}
              \label{Brtp_obs}%
    \end{figure}

We used the synoptic data starting with June 2010 till August 2019, which corresponds to Carrington rotation (CR) 2097 to 2220. Fig. \ref{Brtp_obs} shows a sample of a synoptic vector magnetogram for CR 2103, with B$_\mathrm{r}$ component displayed on top, B$_\mathrm{\theta}$ in the middle and B$_\mathrm{\phi}$ at the bottom.

For verifying the consistency of the boundary data with the force-free assumption, we evaluated the flux, force, and torque for all of the synoptic maps used. While the values obtained for torque are comparable with the one by \cite{2012SoPh..281...37W}, the values for flux and force are in some cases one order of magnitude higher but are still low enough. 



\section{Results}
The magnetic activity is highly influenced by the appearance of the strong concentrations of magnetic flux, which can take the shape of sunspots, faculae, inter-granular high flux concentrations. 

\begin{figure}[ht]
 \centering
  \includegraphics[width=10cm, trim = 40 235 10 225, clip]{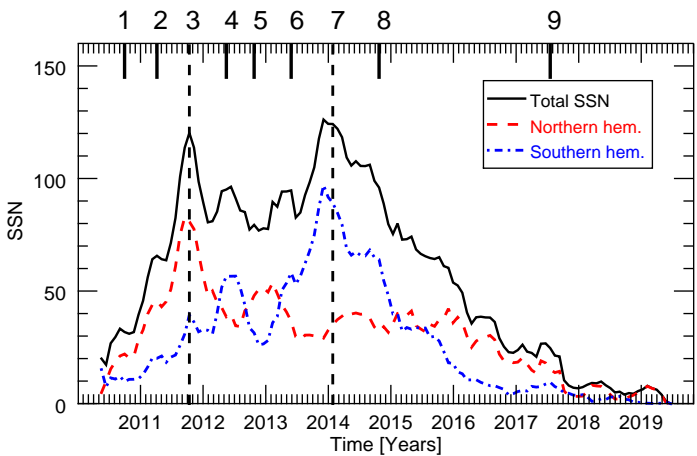}
  \caption{Sunspot number variation during solar cycle 24. Running average over the four CR is shown with a solid black line for the total SSN, with a dashed red line for the NH and with a dash-dotted blue line for the SH. The vertical marks numbered from 1 to 9 shows points of interest described in the text. Data source: WDC-SILSO, ROB, Brussels.}
 \label{SSN}%
\end{figure}

A good proxy for the temporal evolution of a cycle's magnetic activity is represented by the SSN within that cycle. Using the daily total SSN database of World Data Center (WDC) - Sunspot Index and Long term Solar Observations (SILSO), Royal Observatory of Belgium (ROB), we derived the SSN for each CR between June 2010 (CR 2097) and August 2019 (CR 2220). Fig. \ref{SSN} shows the running average over the four CR of the total SSN (solid black line), the northern (dashed red line) and southern (dash-dotted blue line) hemisphere contributions.

For all of the figures, we set points of interest and marked them. Mark number 1 and 2 shows the tip of an oscillatory like pattern during the ascending phase of the cycle. The dashed vertical lines 3 and 7 mark the sunspot cycle peaks, known also as Gnevyshev peaks, identified as the highest values in the SSN data averaged over four CR. Each hemisphere has its Gnevyshev peaks, number 3 and 5 for the NH, and 4 and 7 for the SH. The mark 6 is part of the Gnevyshev gap in the total SSN. Number 8 mark the solar cycle maximum. It is not identifiable as a maximum in the evolution of the SSN (Fig. \ref{SSN}), but it shows a clear maximum in the variation of the magnetic flux at Sun (Fig. \ref{Flux_phot_HMI}, \ref{Flux_hi}, \ref{Ratios_CR}) and Earth (Fig. \ref{Br_1au}), in the free energy (Fig. \ref{Efree}) and flare index (Fig. \ref{Flare_index}). The mark 9 is the last remarkable rise in the cycle activity which is not clearly visible in the SSN variation but is more prominent in the variation of the magnetic flux (Fig. \ref{Flux_phot_HMI}b, \ref{Flux_hi}, \ref{Ratios_CR}).

\begin{figure}[ht]
 \centering
  \includegraphics[width=9.0cm, trim = 0 0 0 0, clip]{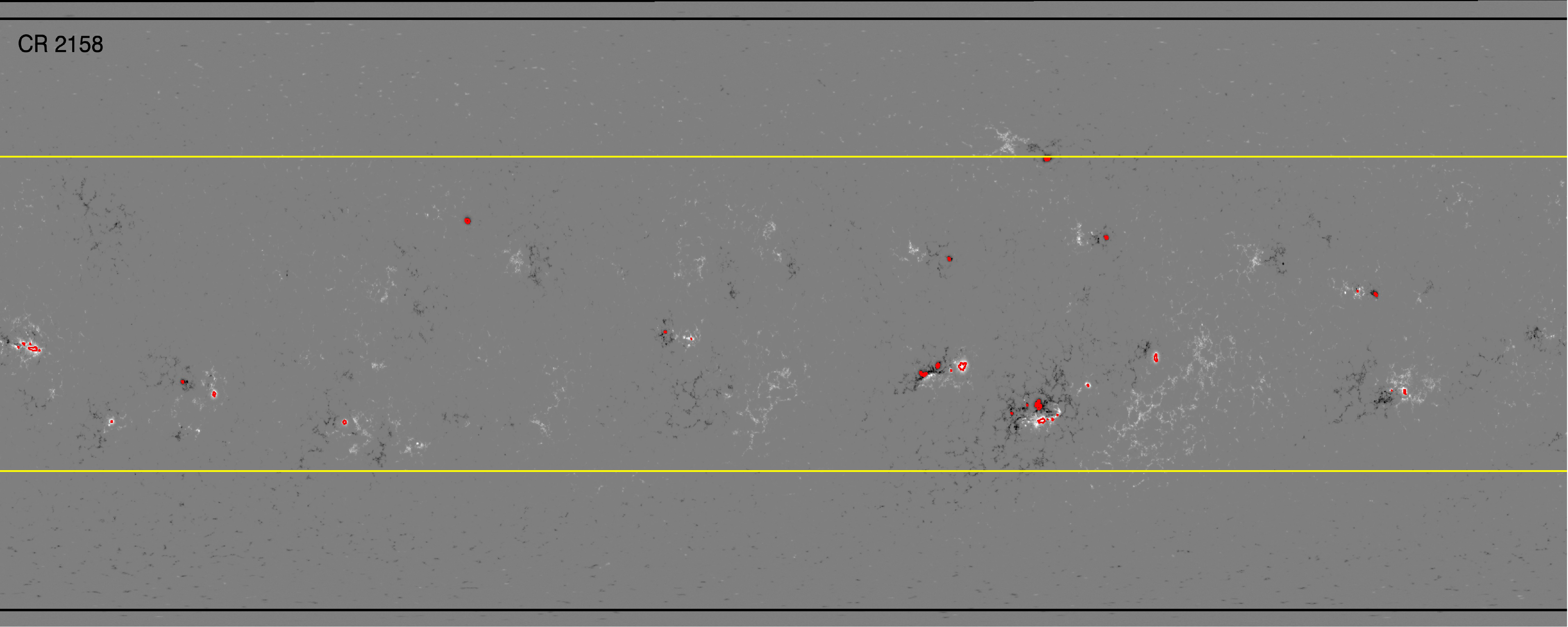}
  \caption{HMI radial magnetic field at the CR 2158. The y axis of the image is scaled with the sine of the latitude. The yellow stripes are drawn at $\pm$30$^\circ$ and the black lines are the demarcations for $\pm$ 70$^\circ$. The red contour represents UPB.}
 \label{HMI_CR_demarc}%
\end{figure}

\begin{figure}[!h]
 \centering
 \includegraphics[width=9.2cm, trim = 45 240 30 235, clip]{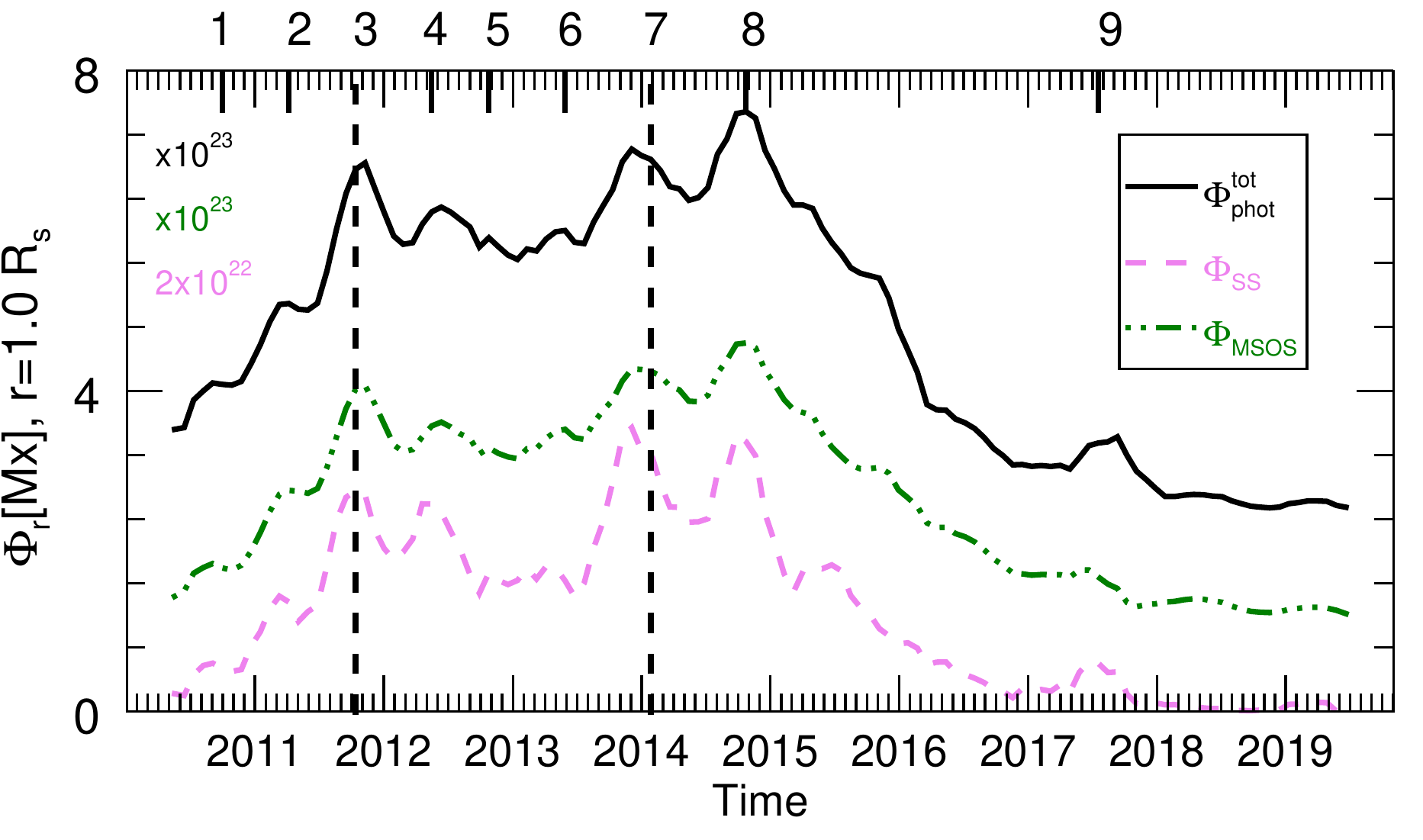}
  \put(-25,75){(a)}
  
   \includegraphics[width=9.2cm, trim = 50 240 30 235, clip]{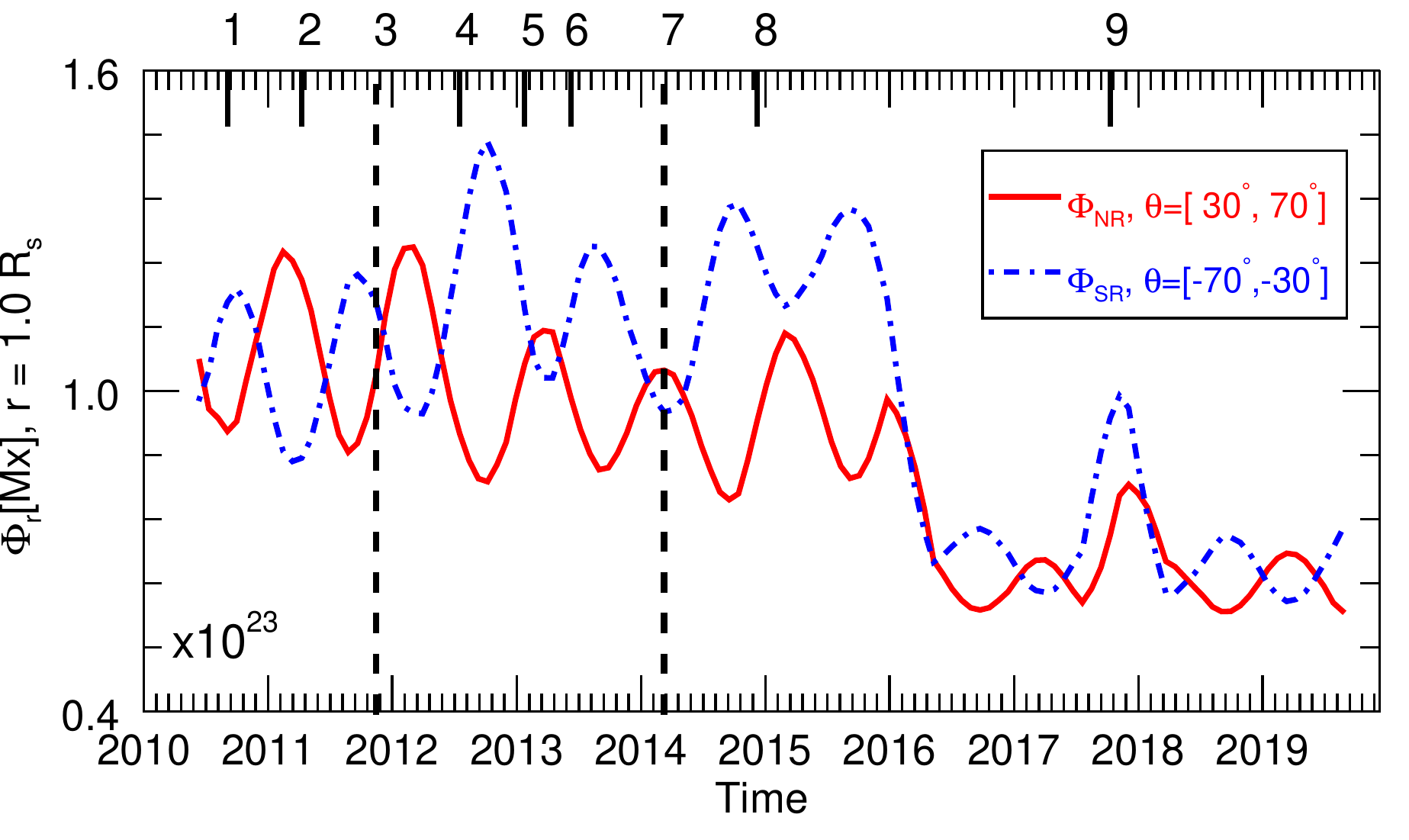}
   \put(-25,75){(b)}
  \caption{Evolution with time of the photospheric magnetic flux calculated from the HMI data. (a) Evolution of the total photospheric magnetic flux ($\Phi^\mathrm{tot}_\mathrm{phot}$ ) in black solid line, sunspot flux ($\Phi_\mathrm{SS}$) in violet dashed line within $\pm$ 30$^\circ$ latitude, flux of the magnetic structures other than sunspots ($\Phi_\mathrm{MSOS}$) within $\pm$ 30$^\circ$ latitude in triple dash-dotted green line}. (b) Evolution of the magnetic flux from the northern $\Phi_\mathrm{NR}$ (solid red line) and southern region $\Phi_\mathrm{SR}$ (dash-dotted blue line).
 \label{Flux_phot_HMI}%
\end{figure}

\begin{figure}[ht]
   \centering
  \includegraphics[width=9.1cm, trim = 42 230 41 213, clip]{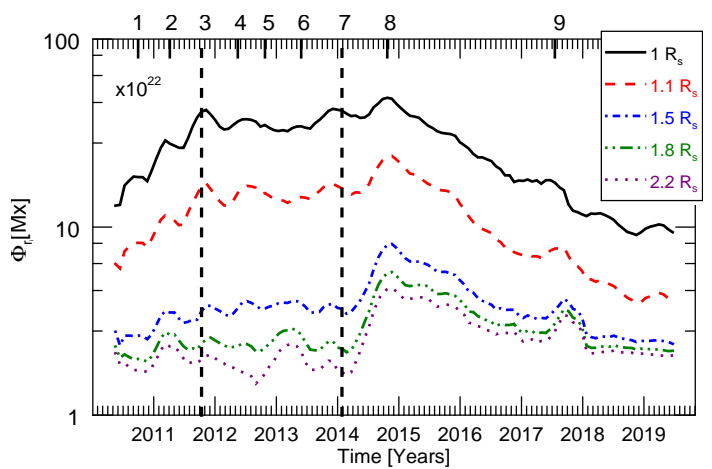}
   \caption{Variation with time of the total unsigned magnetic flux $\Phi_{r_i}$ derived from the solution of the NLFFF extrapolation at r = 1.0 $R_\mathrm{s}$ (black solid line ), 1.1 $R_\mathrm{s}$ (dashed red line), 1.5 $R_\mathrm{s}$ (dash-dotted blue line ), 1.8 $R_\mathrm{s}$ (triple dot-dashed green line), 2.2 $R_\mathrm{s}$ (dotted purple line). The flux is shown on a logarithmic scale. }
              \label{Flux_hi}%
    \end{figure}

\begin{figure*}[ht]
   \centering
    \includegraphics[width=18cm, trim = 30 210 20 210, clip]{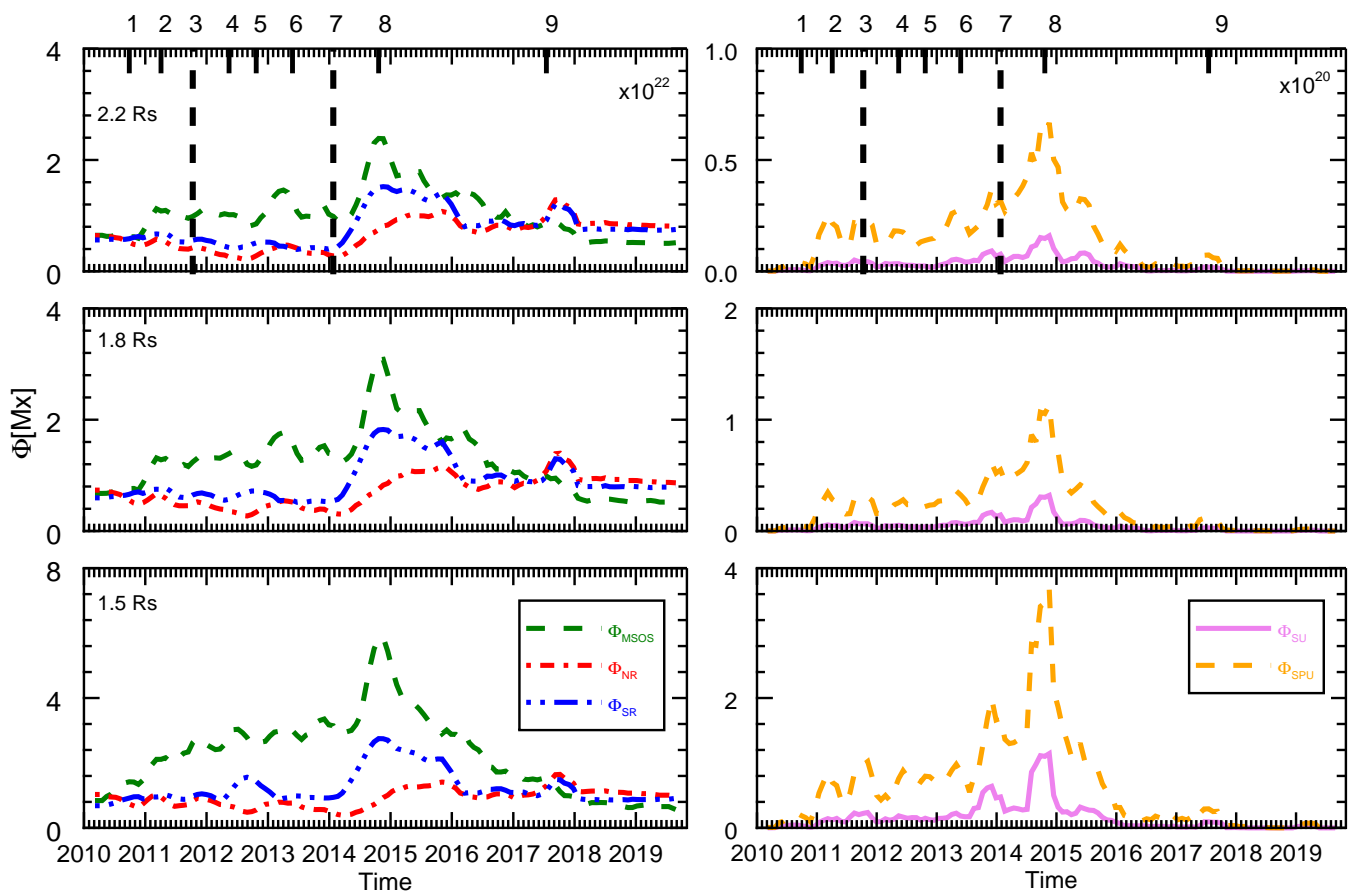}
   \caption{Right panel: Evolution of the unsigned flux derived from the solution of the NLFFF extrapolation for the sunspot umbra (solid violet line) and sunspot penumbra (dashed light orange line) within $\pm$ 30$^\circ$ latitude at heights r = 1.5 (bottom panel), 1.8 (middle panel) and 2.2 (top panel) $R_\mathrm{s}$. Left panel:  Evolution of the unsigned flux derived from the solution of the NLFFF extrapolation for the magnetic structures other than sunspot umbra (MSOSU) within $\pm$ 30$^\circ$ latitude (dashed green line), for the NR $\Phi_\mathrm{NR}$ (red dash-dot line) and the SR $\Phi_\mathrm{SR}$ (triple dot-dashed blue line)} at heights r = 1.5 (bottom panel), 1.8 (middle panel) and 2.2 (top panel) $R_\mathrm{s}$.
    \label{Ratios_CR}%
    \end{figure*}

The variation of the magnetic flux gives more detailed information about the evolution of the cycle's magnetic activity than the variation of the SSNs. Using the HMI radial magnetic field, we calculated the temporal variation of the total unsigned magnetic flux 
\begin{equation}
 \centering
  \Phi_\mathrm{r}=\int^{2\pi}_0 \int^{\theta_\mathrm{max}}_{\theta_\mathrm{min}} |B_\mathrm{r}(r,\theta,\phi)|\, r^2 \sin \theta \,\mathrm{d} \theta \, \mathrm{d} \phi,
  \label{eq_flux}
\end{equation}

in the photosphere (r = 1 $R_\mathrm{s}$) (Fig. \ref{Flux_phot_HMI}) and based on the NLFFF solutions, we derived the total unsigned flux, $\Phi_\mathrm{r}$ (Fig. \ref{Flux_hi}) at various heights above the photosphere (r = 1 ... 2.5 $R_\mathrm{s}$). For the flux calculations we differentiate between five latitudinal regions (Fig. \ref{HMI_CR_demarc}): the southern region (SR) flux ($\Phi_\mathrm{SR}$) with $\theta = $ [-70$^\circ$,-30$^\circ$], the center region with $\theta = $ [-30$^\circ$,30$^\circ$] which is the sunspot latitudinal band during a cycle \citep{2020LRSP...17....4C}, the northern region (NR) flux ($\Phi_\mathrm{NR}$) with $\theta= $[30$^\circ$,70$^\circ$] and northern ($\Phi_\mathrm{NH}$) and southern hemispheres ($\Phi_\mathrm{SH}$) with $\theta = [\pm$ 70$^\circ$,0$^\circ$]. We separate the center region in sunspot umbra flux ($\Phi_\mathrm{SU}$), sunspot penumbra flux ($\Phi_\mathrm{SPU}$) and flux originating from magnetic structures other than sunspots ($\Phi_\mathrm{MSOS}$). We choose the latitudinal limit of $\pm 70$ for comparison with the NLFFF extrapolation.

For calculating the total unsigned magnetic sunspot umbra flux $\Phi_\mathrm{SU}$, we considered all the photospheric flux larger than 1693 G. We choose this threshold based on previous empirical studies. Using HINODE data, \cite{2018A&A...611L...4J} found the sunspot umbra-penumbra boundary (UPB) has the most probable value of 1867 G. \cite{2019ApJ...873L..10M} points out that \cite{2018A&A...620A.104S}, using data from HMI/SDO, reports the value of 1693 G as UPB. Since we use HMI/SDO data, we chose the value reported by \cite{2018A&A...620A.104S}. As there is no clear separation between the penumbra and the quiet sun, we use the value from \cite{2003A&A...411..257S} as a threshold, which says that the magnetic field reaches 800 G at the outer sunspot boundary. We consider the total unsigned magnetic sunspot penumbra flux ($\Phi_\mathrm{SPU}$) the flux between 800 G and 1693 G.

The unsigned magnetic flux within $\theta = [-30^\circ,30^\circ]$ shows the closest resemblance to the SSN variation, mainly the flux from the sunspots $\Phi_\mathrm{SS}$, defined as the sum of the flux from the sunspot umbra ($\Phi_\mathrm{SU}$) and penumbra ($\Phi_\mathrm{SPU}$) (Fig. \ref{Flux_phot_HMI} panel (a), violet dash-dotted curve). A difference with the SSN evolution is seen in the total magnetic flux $\Phi^\mathrm{tot}_\mathrm{phot}$ and $\mathbf{\Phi_\mathrm{MSOS}}$ at the mark labelled with number 8, which is about six months after the second Gnevyshev peak (mark number 7), that is, at the time when SSN are already in their descending phase. The maximum of the cycle activity in terms of emerging flux occurs at the end of 2014, a different epoch than the SSN maximum. For the entire solar cycle, $\mathbf{\Phi_\mathrm{MSOS}}$ was the main contributor to the total flux in the photosphere. The value of the total unsigned flux from sunspots $\Phi_\mathrm{SS}$ is one order of magnitude lower than the $\mathbf{\Phi_\mathrm{MSOS}}$ and a factor of two to five lower than the value from N-S regions. A similar result was obtained by  \cite{2019RAA....19...69J} who reported that the total flux from the inter-network magnetic field is larger than from the sunspot/pores over the solar cycle 24. They used a threshold of 950 G and 1.0x10$^{20}$ \mbox{Mx} to differentiate between sunspot/pore and the rest of the magnetic structures. 
According to \cite{2018Ge&Ae..58.1159A} and \cite{2019A&A...629A..45N} the ARs variability and total unsigned flux present a good correlation with the SSN variation for the cycle 24. In our results, the total unsigned flux variation does not follow the SSN variation. The source of the unsigned flux at mark number 8 has its roots not only in the AR flux but also in the MSOS flux.

he unsigned magnetic flux from the northern and southern regions presented in Fig. \ref{Flux_phot_HMI}b manifest an anti-phase relationship during most of the cycle excepting the period around mark number 9, when the hemispherical regions are almost in phase. The temporal synchronisation has an impact on the $\Phi_\mathrm{r}$ (Eq. \ref{eq_flux}) at each atmospheric level from the photosphere (Fig. \ref{Flux_phot_HMI}a ) to corona (Fig. \ref{Flux_hi}). 

The beginning of 2016 marks an event in two steps for the evolution of the hemispherical $\Phi_\mathrm{r}$. The first part of the event is visible as a sudden drop in $\Phi_\mathrm{r}$ from the northern and southern region (Fig. \ref{Flux_phot_HMI}b), seen also in the variation of the total unsigned flux (Fig. \ref{Flux_phot_HMI}a). The second part of the event is the beginning of a lower-level $\Phi_\mathrm{r}$ in respect to the earlier phase of the cycle. From observations, the magnetic flux from the sunspot band diffuses and migrates towards the poles by meridional circulation \citep{1989Sci...245..712W}. A possible explanation for the first part of the event is that during the migration period, a larger part of the flux cancelled and the effect was seen in the unsigned magnetic flux which suffered a sudden drop. A stronger decrease in the SH as observed in Fig. \ref{Flux_phot_HMI}b suggest that the sink or cancellation of the magnetic field was more efficient than in the NH. 

The second part of the event can probably be explained with the dynamo theory. An important ingredient in the solar dynamo models is the transport of the flux from the mid to high latitudes. Some models consider single or multiple-cell meridional circulation which transports the flux from the sunspot band towards the poles. The presence of turbulence gradients in convectively unstable layers give rise to the turbulent pumping mechanism \citep{2008A&A...485..267G} which transports the magnetic fields from (near-) surface solar layers to the deep interior \citep{1998ApJ...502L.177T, 2001A&A...365..562D, 2003A&A...401..433Z,2016ApJ...832....9H}. The downward turbulent pumping is considered to be a better mechanism for reducing the polar field strength \citep{2020LRSP...17....4C}. From observations of cycle 23, \cite{2010Sci...327.1350H} showed that the meridional flow velocity is higher at solar minimum than SSN maximum. \cite{2020LRSP...17....2P} mentions that the overall flow at mid-latitudes is slower before and during maxima, and faster during the decay phase. A faster transport at minimum from the sunspot band to higher latitudes would mean an increase in the unsigned flux above the sunspot band boundary. Cycle 24 observations shows that in the descending phase of the cycle there is a decrease in the flux strength within latitudinal interval $\pm$30 to $\pm$70 (Fig. \ref{Flux_phot_HMI}b). To explain the second part of the event, turbulent pumping might be a solution as considered by \cite{2016ApJ...832....9H}. However, we believe it should happen not only at the poles as suggested by \cite{2020LRSP...17....4C}. \cite{2010ASSP...19...86N} considers turbulent pumping the most dominant flux transport mechanism for downward transport of the poloidal field. Maybe another mechanism for explaining the second part of the event is the appearance of multiple meridional circulation cells during the descending phase of the cycle. This would effect the subduction of more magnetic field compared with the previous cycle phase when only one meridional cell would activate. 
 
 Over the cycle, the unsigned magnetic flux changes its behaviour with coronal altitudes in respect to the photosphere (Fig. \ref{Flux_hi}). 
    Around the mark labelled with number 8, the flux $\Phi_{\mathrm{r}_i}$ at heights above 1.5 $R_\mathrm{s}$ shows an increase by almost a factor of two, much more pronounced than  $\Phi_{\mathrm{r}_i}$ at lower heights. The reason for this increase might be due to a concentrated strong and complex magnetic field in the photosphere, which has a strong impact on higher layers of the atmosphere.
    
    In Fig. \ref{Ratios_CR} we show the temporal evolution at r = 1.5, 1.8, 2.2 $R_\mathrm{s}$ of the unsigned $\Phi_\mathrm{r}$ from sunspots (right column), MSOS, northern and southern regions (left column). Except for the NR, all the other components have their highest flux value at mark number 8. 
    
From the beginning of 2016 till the end of the cycle, the unsigned flux from northern and southern regions have the same contribution to the $\Phi_\mathrm{r}$. Starting with the middle of 2017 (Fig. \ref{Ratios_CR}, left column), the hemispherical region flux ($\Phi_\mathrm{NR}$, $\Phi_\mathrm{SR}$) is higher than the $\Phi_\mathrm{MSOS}$ flux from the sunspot band.

On average, over the entire cycle for latitudes between $\pm 30^\circ$ the MSOS flux represents around 56\% of the total flux in the photosphere and the sunspot flux is around 5\%. 

The cycle 24 had more sunspots in the NH than the SH, but stronger flux in the SH. The NH represents 52 \% of the total SSN in Fig. \ref{SSN}. On the other hand, the unsigned flux at the photosphere (Fig. \ref{Flux_phot_HMI}b) and corona (Fig. \ref{Ratios_CR}, left column) is dominated by the SR with approximately 53 \%.

\begin{figure}[ht]
   \centering
\includegraphics[width=9.1cm, trim = 48 245 40 240, clip]{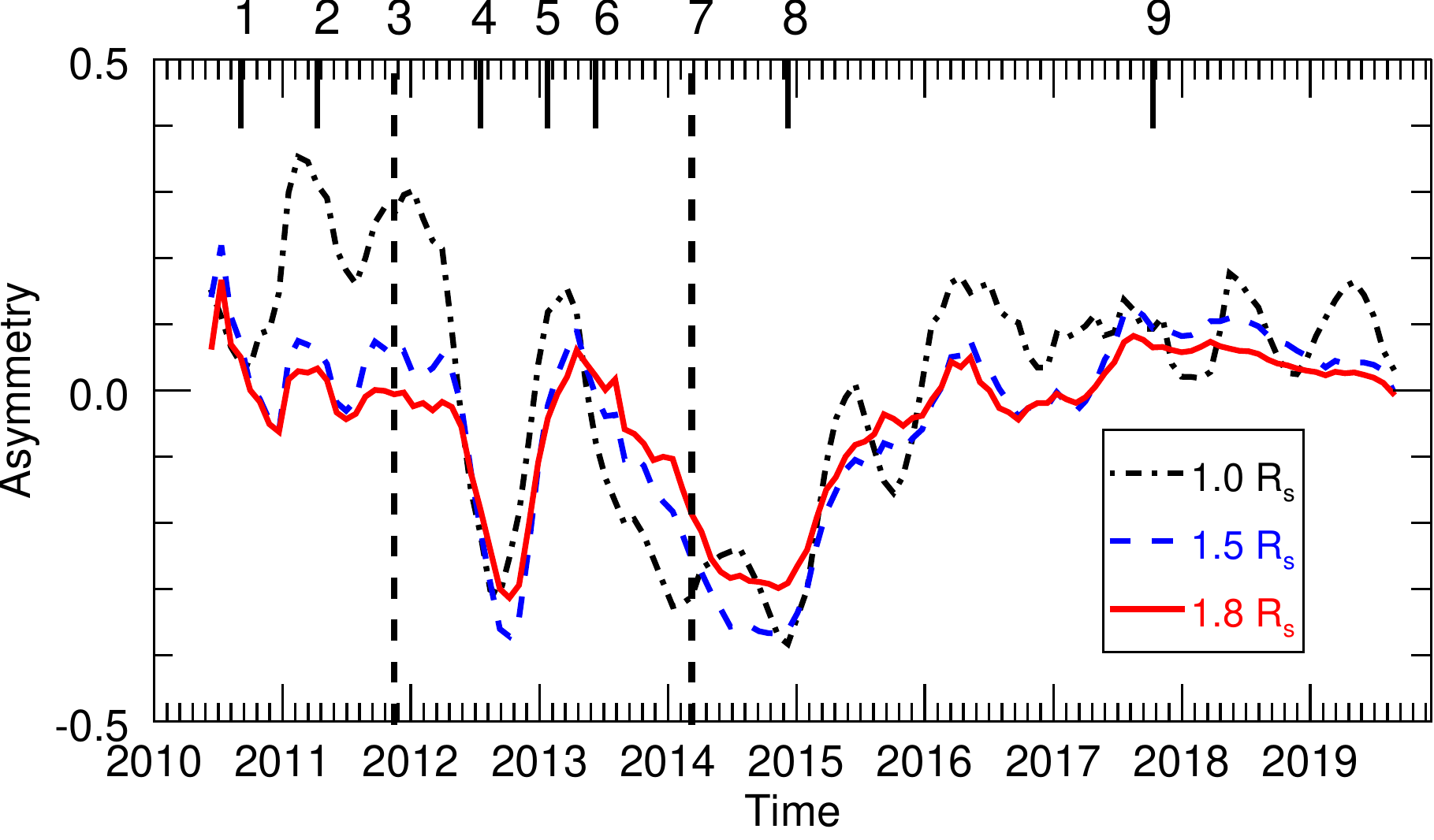}  
 \put(-220,30){(a)}\\
\includegraphics[width=9.1cm, trim = 90 285 49 240, clip]{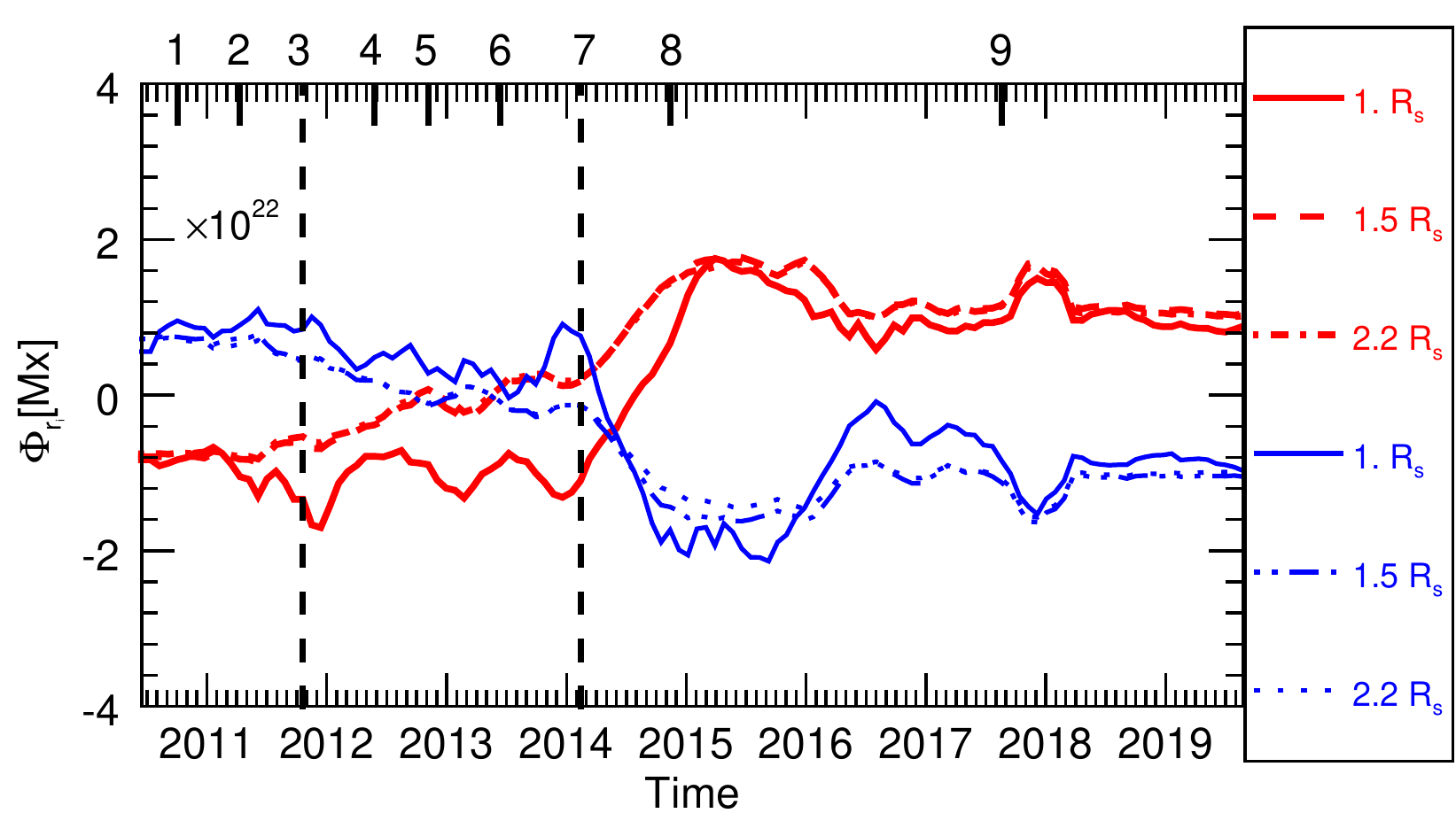} 
 \put(-230,40){(b)}

   \caption{Panel (a): Normalised asymmetry flux index for the northern and southern hemisphere, $\theta$ = [0$^\circ$, $\pm70^\circ$] derived from the solution of the NLFFF extrapolation at r = 1.0 Rs (dash-dotted black line), 1.5 $R_\mathrm{s}$ (dashed blue line) and 1.8 $R_\mathrm{s}$ (solid red line). Panel (b): Signed flux for the NH at r = 1 Rs (thick solid red line), r = 1.5 $R_\mathrm{s}$ (dashed red line), r = 2.2 $R_\mathrm{s}$ (dash-dotted red line) and for the SH at r = 1 $R_\mathrm{s}$ (thin solid blue line), r = 1.5 $R_\mathrm{s}$ (triple dot-dashed blue line), r = 2.2 $R_\mathrm{s}$ (dotted blue line).}
    \label{Asymmetry}%
    \end{figure}

In Fig. \ref{Asymmetry}a we show the normalised asymmetry index for the unsigned magnetic flux at different heights $(\Phi_\mathrm{NH}-\Phi_\mathrm{SH})/(\Phi_\mathrm{NH}+\Phi_\mathrm{SH})$ and in Fig. \ref{Asymmetry}b we plot the evolution of the signed magnetic flux from the northern and SH at r = 1.0, 1.5, 1.8 Rs.

Fig. \ref{Asymmetry}a shows, at all heights, the NH flux contributed to the total flux mainly in the ascending phase, first peak (mark number 3) and the second part of the descending phase of the cycle. A small northern flux dominance during the cycle is also noticeable during the Gnevyshev gap (mark number 5). The year 2016 marks the beginning of a more symmetric N-S flux for the descending phase of the cycle. 

Investigating the SSNs from northern and southern hemispheres of solar cycles 18-23, \cite{2006A&A...447..735T} conclude that the hemispheres are in phase during the solar minimum, while high asymmetries beyond statistical uncertainties are noted in the SSN at maximum. The authors report other studies which showed that the N-S asymmetry is higher at solar minimum. From our data, we observe that the N-S asymmetry of the unsigned flux changes with height. In Fig. \ref{Asymmetry}a, strong asymmetries are identified at mark numbers 2, 3, 4-5, 7 and 8 for r = 1.0 $R_\mathrm{s}$, and mark numbers 4-5 and 8 for r = 1.5, 1.8 $R_\mathrm{s}$. Starting with 1.5 $R_\mathrm{s}$ the N-S contribution becomes more symmetric in regions with initial NH dominance. A possible explanation is that while in the photosphere the northern emergent flux had its source on larger areas (more sunspots), it was weaker than the southern one, so it could not penetrate to the high-layers of the atmosphere as the southern flux. One of the strong asymmetries (between marks 4 and 5) is co-temporal with an equal number of sunspots from the two hemispheres (Fig. \ref{SSN}) and with the highest amplitude in the SR flux (Fig. \ref{Flux_phot_HMI}b). Over all of the atmospheric layers, the strongest asymmetries in the unsigned flux (fig. \ref{Asymmetry}a) occur during the Gnevyshev gap (between marks 4 and 5) and after the SSN maximum (mark number 8). 

The hemispheric asymmetry is also an indication of the hemispheric coupling \citep{1990ApJ...360..296A}. Analysing the N-S asymmetry for the solar cycles 18-23, \cite{2006A&A...447..735T} concluded that the two hemispheres are weakly coupled. According to \cite{1990ApJ...360..296A}, the time-reversal of the polar fields, the weak interdependence of the magnetic field systems, the level of activity which differs significantly and the asymmetry between the rotation rates from the two hemispheres shows a weak coupling between the northern and southern hemispheres.

Evaluating the total unsigned magnetic flux (see Fig. \ref{Flux_phot_HMI}b), we notice an anti-phase relationship in the photosphere between the two hemispheric regions. At heights above 1.4 $R_\mathrm{s}$ (Fig. \ref{Ratios_CR}, \ref{Asymmetry}b) the N-S relation is changing with height, and in the top part of the corona, we find that the N-S regions have almost the same level of activity during most of the cycle. From our results (see Fig. \ref{Ratios_CR}, \ref{Asymmetry}a, b), considering \cite{1990ApJ...360..296A} perspective on coupling, we observe that starting with 2016 the two hemispheres become more coupled compared with the previous part of the cycle, and also the coupling between the hemisphere increases with height (Fig. \ref{Ratios_CR}, \ref{Asymmetry}a, b). The correlation between the unsigned flux from the N-S hemispheres is increasing with height from approximately 0.55 to 0.85.

The observation of the polarity inversion is used as a constraint in the models to predict the evolution and strength of the following solar cycle. In Fig. \ref{Asymmetry}b, we plot the evolution of the magnetic signed flux from the northern and southern hemispheres at three heights in the atmosphere.
\cite{2019SoPh..294..137P}, in a study on the photospheric polar ($\pm[50,90]$) magnetic field reversal for the last three cycles, reports for the cycle 24 multiple reversals in the  NH and single SH reversal. They also report other studies which identify single and multiple reversals in the NH and single reversal in the SH. From our data, for the photospheric magnetic field within $\pm[45,90]$ we note multiple polarity changes for both hemispheres, while for the one within$\pm[70,90]$ we note single polarity changes in both hemispheres. At heights above 1.5 $R_\mathrm{s}$, the flux (Fig. \ref{Asymmetry}b ) has more sign changes than in the photosphere.

\cite{2019SoPh..294..137P} finds that the NH completes the reversal in April 2014 and the SH in April-May 2015. \cite{2015ApJ...798..114S} shows the averaged magnetic field between $\pm[60,90]$ reversed sign in November 2012 (N hem) and in March 2014 (S hem). \cite{2018A&A...618A.148J} reports that the SH reversed polarity during the mid-2013, while the NH started the field reversal as early as June 2012 and completed around November 2014.  According to our results, the first change of polarity in the NH (r = 1 $R_\mathrm{s}$) occurs in September-October 2014 (Fig. \ref{Asymmetry}b, solid red line). The SH changes sign in March-April 2014 in agreement with \cite{2015ApJ...798..114S}, epoch close to the second cycle peak. \cite{2018A&A...618A.148J} reports a study which finds different polarity reversal times based on observations made in the low solar corona (17 GHz microwave emission). From our results, we can confirm that the polarity change can vary also with height and not only with time (Fig. \ref{Asymmetry}b). The reason might be that the differential rotation occurs not only with latitude but also with height and time \citep{2001ApJ...548L..87V, 2010NewA...15..135B}.
The year 2016 marks the beginning of simultaneous polarity reversals at all atmospheric layers (Fig. \ref{Asymmetry}b) until the end of the cycle.


\subsection{Magnetic energy}

Most dynamic processes in the corona are driven by the Lorenz forces. Since the potential field for a given vertical surface flux distribution in the photosphere has the lowest possible energy, a field configuration involving currents must have an excess energy, the so-called free energy, if it matches the same photospheric vertical flux distribution. The free energy can therefore serve as a proxy for the probability and the strength of dynamic processes in the corona, such as flares, prominence eruption, or coronal mass ejections (CMEs). \cite{2008A&A...484..495T} showed that free energy increases right before a flare eruption and it decreases again after the eruption.

We estimate the free energy, 
\begin{equation}
 \centering
  E_\mathrm{free} = \frac{1}{8\pi} \int^\mathrm{2.5R_s}_\mathrm{1.0R_s} \int^{2\pi}_0 \int^{\theta_\mathrm{max}}_{\theta_\mathrm{min}} (\Vec{B}_\mathrm{NLFFF}^2 - \Vec{B}_\mathrm{pot}^2)\, r^2 \sin \theta \,\mathrm{d} \theta \, \mathrm{d} \phi\, \mathrm{d} r,
  \label{eq_energy}
\end{equation}
 by calculating the actual coronal field $\Vec{B}_\mathrm{NLFFF}$ as NLFFF extrapolation and determine the potential $\Vec{B}_\mathrm{pot}$ using the same vertical surface flux distribution in the photosphere.

\begin{figure}[ht]
		\hspace{-0.3cm}\includegraphics[width=10cm, trim = 20 245 0 240, clip]{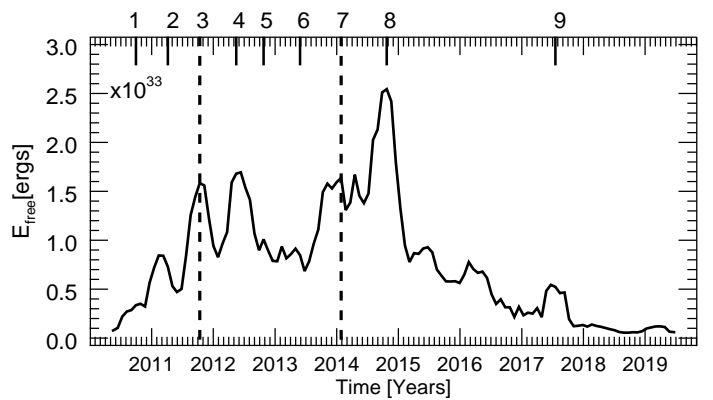}
	\caption{The variation with time of the magnetic free energy obtained from the evaluation of the NLFFF magnetic flux.}
	\label{Efree}%
\end{figure}

In Fig. \ref{Efree} we show the time evolution of the free energy obtained in the latitudinal range $\pm$ 70$^\circ$. Compared to the SSN variation (Fig. \ref{SSN}), we find enhancement at the first major peak at mark label 3 (dominated by the northern sunspot) and the first minor peak (dominated by the southern sunspot), however the latter with much higher energy. During the second half of the activity cycle, the correlation between SSN and the free energy is worse. The energy variation shows similarities to the unsigned magnetic flux variation in the corona (Fig. \ref{Flux_hi}), emphasizing the peak at mark number 8.

\begin{figure}[ht]
	\centering
		\includegraphics[width=9.1cm, trim = 50 220 49 225, clip]{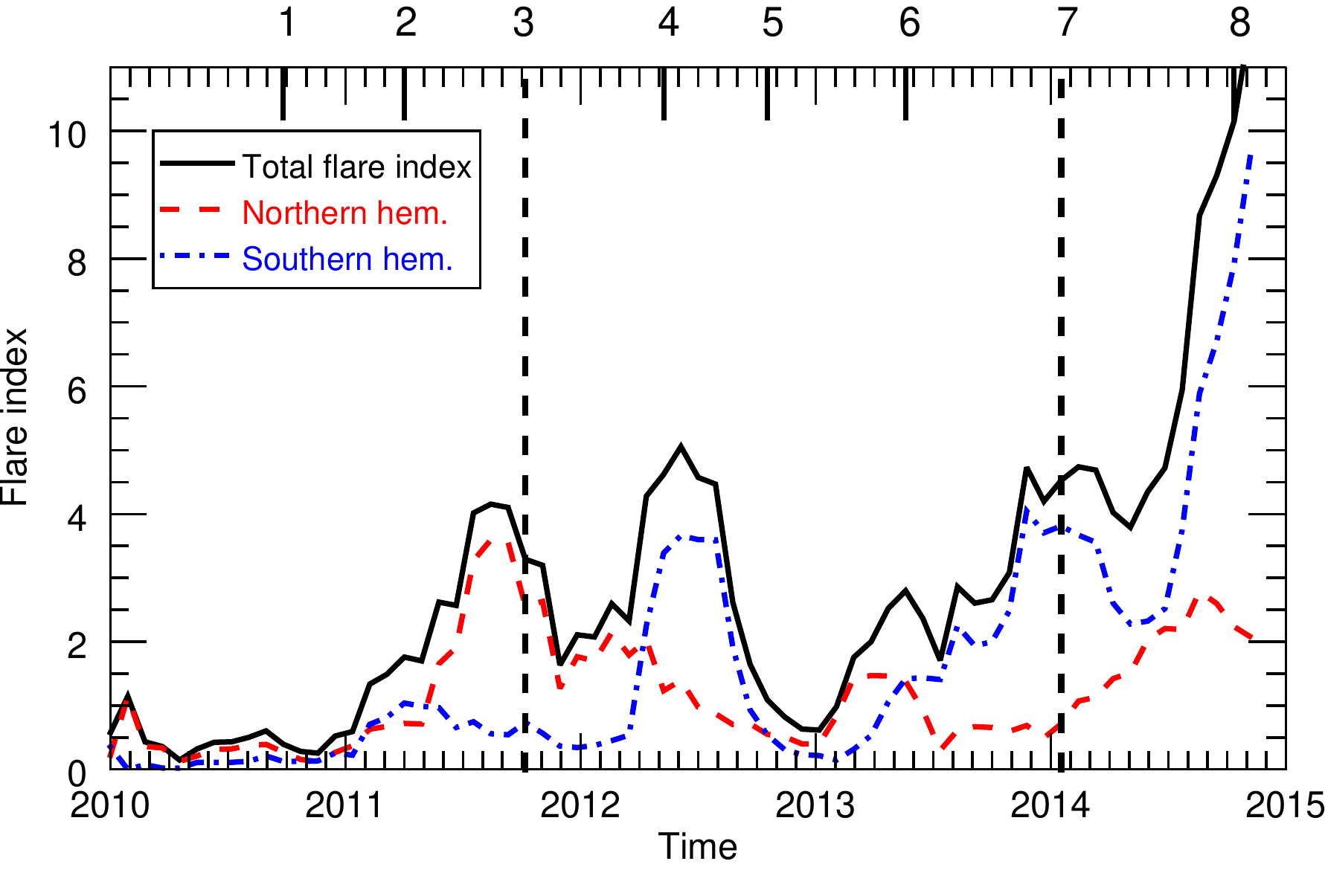}
 \put(-220,100){(a)}
 
		\includegraphics[width=9.1cm, trim = 40 230 20 220, clip]{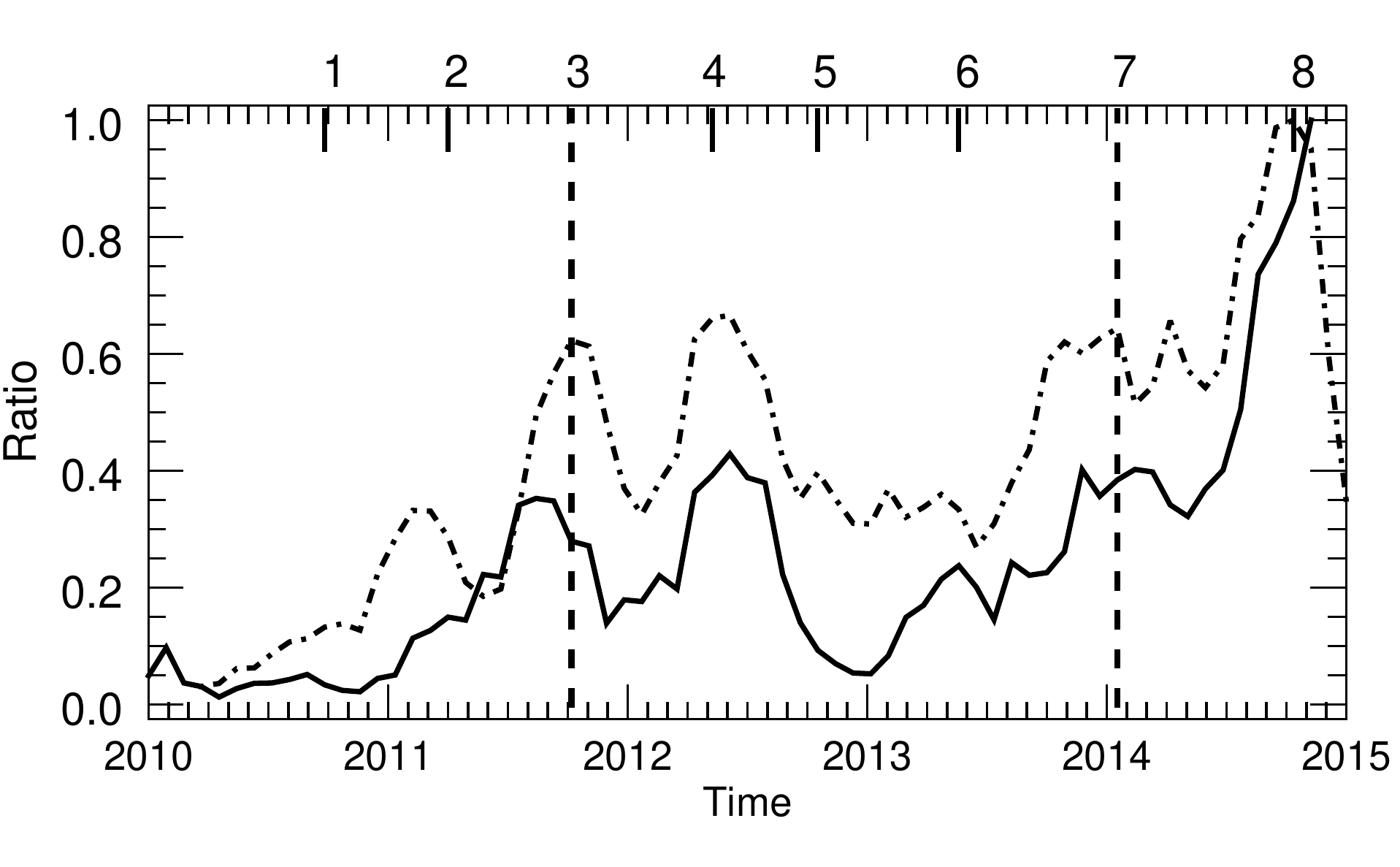}
		\put(-220,120){(b)} 
	\caption{Panel (a): Variation of the flare index during January 2010 and December 2014: total flare index (solid black line), the contribution from the NH (dashed red line) and from the SH (dash-dotted blue line). Panel (b): Ratio of the total flare index to the maximum of it (solid black line) and the ratio of the free energy to the maximum of it (dash-dotted black line).}
	\label{Flare_index}%
\end{figure}

In Fig. \ref{Flare_index}a we display the four-CR running average flare index published by the Kandilli Observatory and Earthquake Research Institute at the Bogazici University and made available through the National Oceanic and Atmospheric Administration (NOAA), National Geophysical Data Center (NGDC)\footnote{ \url{https://www.ngdc.noaa.gov/stp/space-weather/solar-data/solar-features/solar-flares/index/flare-index/}}. With the solid black line we show the total flare index. The contribution from the NH we plot with a dashed red line and the contribution from the SH we plot with a dash-dotted blue line. The flare index is roughly proportional to the mean energy emitted by flares, estimated by the product of importance of the flare \citep[see][for details]{2003SoPh..214..375O} and its duration in minutes. The importance takes into account the area and the intensity of the flare\footnote{\url{https://previ.obspm.fr/index.php?page=flares\&sub=qbsa}}.

In Fig. \ref{Flare_index}b we show the ratios of total flare index to the maximum of the total flare index (solid black line), of free energy to the maximum of the free energy (dash-dotted black line). We calculated the match between free energy and flare index using Pearson's correlation coefficient, which gave us a value of 0.88 for the four-CR running averaged curves. 

\begin{figure}[ht]
	\centering
          \includegraphics[width=9.2cm, trim = 45 80 20 50, clip]{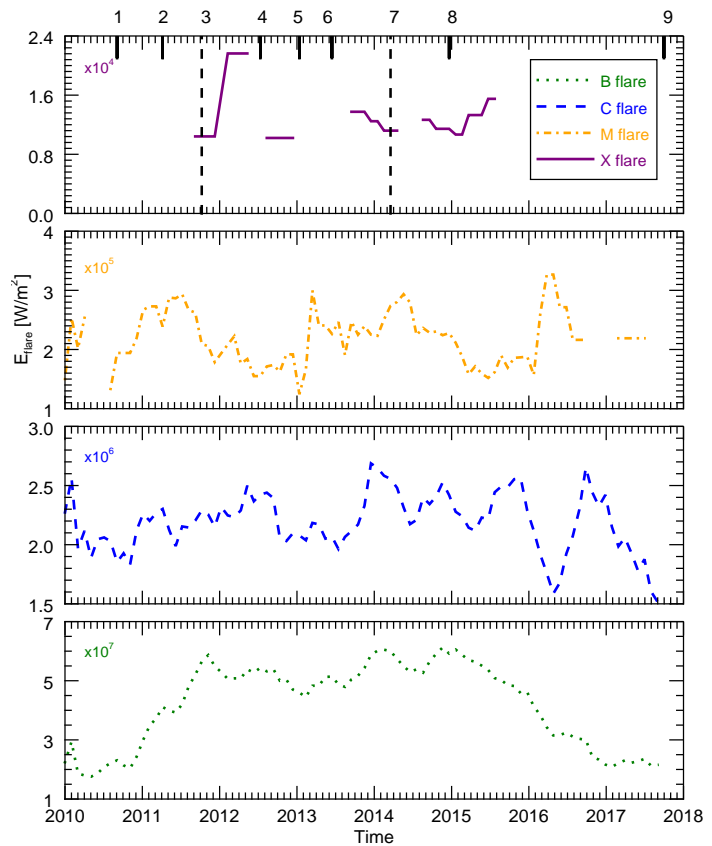}
	\caption{Evolution of the flare energy for the B class (dotted green line), C class (dashed blue line), M class (dash-dotted orange line), X class (solid purple line) flares. The energy is shown on a logarithmic scale.}
	\label{Flares}%
\end{figure}

In Fig. \ref{Flares} we show the four-CR running average of the flare energy. The green dotted curves represents the averages for B class flares while the dashed blue, dash-dotted orange, solid  purple curves shows the C, M, X class flares. The figure is constructed based on the data from the Geostationary Operational Environmental Satellite (GOES) satellite through the NOAA NGDC\footnote{ \url{https://www.ngdc.noaa.gov/stp/solar/solarflares.html}}.  

The X-ray flares are classified according to the peak burst intensity measured at the earth in the 1 to 8 angstrom band. There are four classes: the B class flare has an energy peak less than 10$^{-6}$ $\mbox{W m}^{-2}$, the C class has the energy between 10$^{-6}$ and 10$^{-5}$ $\mbox{W m}^{-2}$, M class between 10$^{-5}$ and 10$^{-4}$ $\mbox{W m}^{-2}$ and X class flares between 10$^{-4}$ and 10$^{-3}$ $\mbox{W m}^{-2}$.

In the flare index variation, we identify a first spike shortly before the first solar cycle peak (mark number 3). The free energy (Fig. \ref{Flare_index}b, dash-dot line) is co-temporal with the first Gnevyshev peak (Fig. \ref{Flare_index}b, mark number 3) and not perfectly aligned with the first spike of the flare index (Fig. \ref{Flare_index}b, solid line). We believe that the free energy peak after the flare index is not the cause of the latter. Previous studies \citep{2008A&A...484..495T} proved an increase of free energy before an AR flare occurrence and a free energy decrease after the eruption. One reason might be that the dependence of the flaring time which is taken into account when calculating the flare index is influencing the time variation. While the B class flares follows the unsigned flux variation including even the oscillatory pattern, the behaviour is not present for the C, M, X class flares. The M class flares show a first energy peak right before mark number 3. These M flares are responsible for the first flare index spike which appears right before the first cycle peak. We observe the appearance of the X class flares during all SSN peaks, which are prominent in the magnetic flux and energy: the two major peaks, mark numbers 3 and 7, the first (southern) minor peak, mark number 4. In addition, a large number of X class flares are present in near peak mark number 8 which only shows in magnetic flux and energy. While the strongest flare occurs early in the activity cycle at the first major peak, the highest occurrence rate is at peak mark number 8 during the descending phase of the SSN.

\subsection{Magnetic field at Earth}

The estimation of the magnetic field at Earth is based on the open modelled magnetic field in the solar corona at the source surface 

\begin{equation}
 \centering
 |B_\mathrm{r_{1AU}}| = \frac{\Phi_\mathrm{open}}{4\pi r^2_\mathrm{1au}}. 
  \label{Br_1au_eq}
\end{equation}

The height at which the coronal field is considered to be "open" is at r = 2.5 $R_\mathrm{s}$, the approximate boundary where the plasma pressure is dominating again over the magnetic pressure.

Based on the potential field and NLFFF extrapolations, we derived the radial magnetic flux at 1 AU (Eq. \ref{Br_1au_eq}). 

\begin{figure}[ht]
	\centering
		\includegraphics[width=9.2cm, trim = 60 250 30 240, clip]{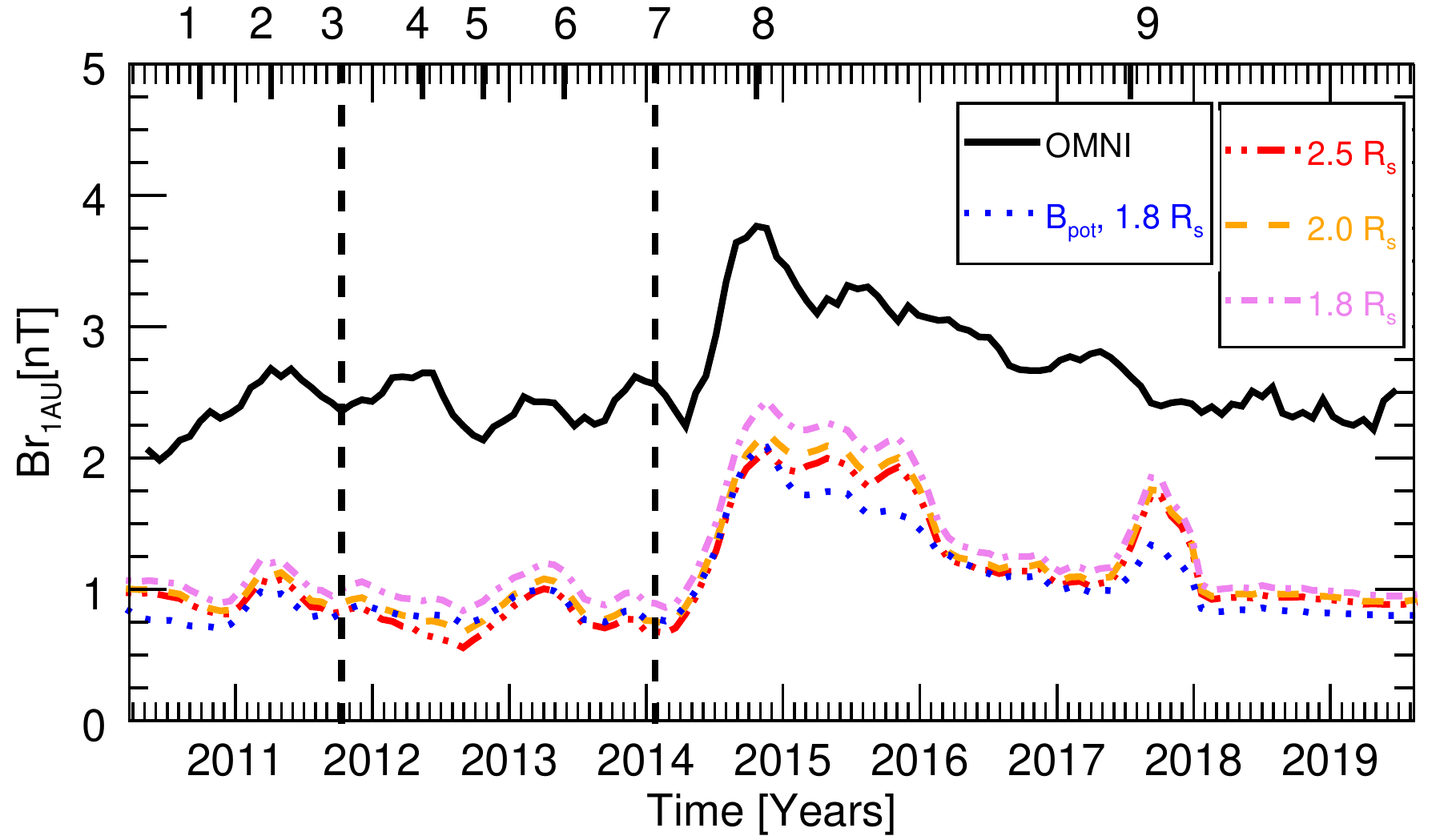}
		\put(-215,110){(a)}
		
		\includegraphics[width=9.2cm, trim = 60 240 30 230, clip]{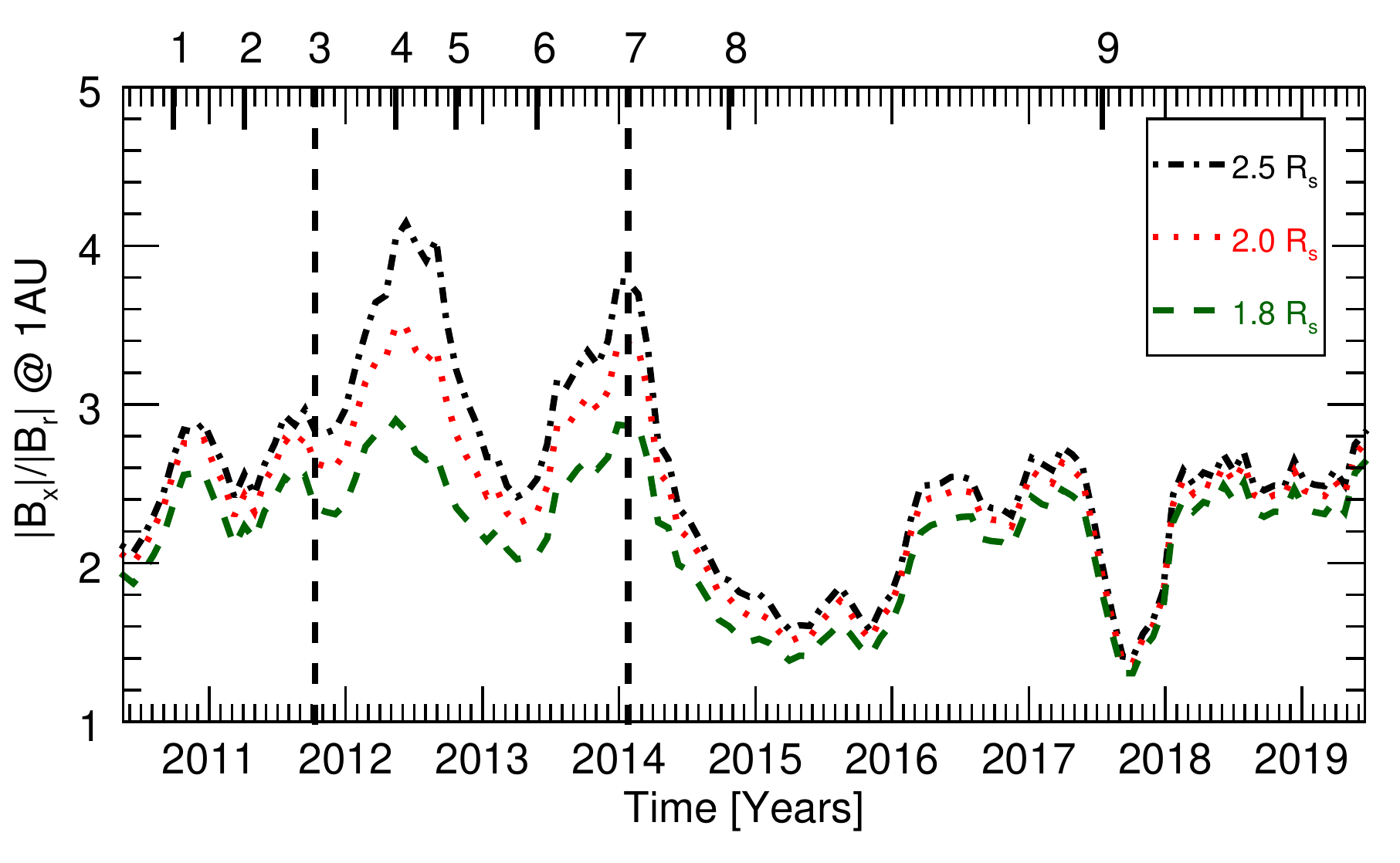}
		\put(-215,110){(b)} 
	\caption{Panel a: Temporal variation of the observed magnetic field (solid black line) at 1 AU together with the magnetic field derived from the solution of the NLFFF extrapolation at height of r = 2.5 $R_\mathrm{s}$ (triple dot-dashed red line), r = 2.0 $R_\mathrm{s}$ (dashed orange line) and r = 1.8 $R_\mathrm{s}$ (dot-dashed violet line) and with the magnetic field derived from the potential field at r = 1.8 $R_\mathrm{s}$ (dotted blue line). Panel b: Ratio between the observed and derived radial unsigned magnetic field at 1 AU. The derived magnetic field was calculated from the solution of the NLFFF at coronal heights of r = 2.5 $R_\mathrm{s}$ (dash-dotted black line), r = 2.0 $R_\mathrm{s}$ (dotted red line) and r = 1.8 $R_\mathrm{s}$ (dashed green line)}.
	\label{Br_1au}%
\end{figure}

In Fig. \ref{Br_1au}a we show the radial component of the observed interplanetary magnetic field at Earth (solid black line), the calculated magnetic field at 1 AU from the NLFFF extrapolation at r = 2.5 $R_\mathrm{s}$ (triple dot-dashed red line), r = 2.0 $R_\mathrm{s}$ (dashed orange line), r = 1.8 $R_\mathrm{s}$ (dot-dashed violet line). The observed field is from OMNIWeb\footnote{\url{https://omniweb.gsfc.nasa.gov/html/ow\_data.html}} data center which is a compilation of data from spacecraft in geocentric or Lagrange one (L1) orbit.
In Fig. \ref{Br_1au}b we show the ratio between the observed field |Bx| at Earth and the one derived with the NLFFF method (|B$_{r1au}$|).

\cite{2017ApJ...848...70L} considers the existence of an "open flux problem" because the calculated flux at Earth based on the open magnetic flux most often obtained from potential field source surface (PFSS) method is much lower than observed flux as in Fig. \ref{Br_1au}a. Using as bottom boundary different magnetograms for the CR 2098, \cite{2017ApJ...848...70L} modelled the magnetic field in the solar corona using potential field source surface (PFSS) and magnetohydrodynamic (MHD) methods from which they calculated the magnetic field at Earth. From the solutions of the two methods, they obtained the magnetic flux at different heights (r = 1.3, 1.4, 2.0, 2.5 $R_\mathrm{s}$) in the atmosphere from which they calculated the magnetic field at one astronomical unit (1 AU). With one exception (r = 1.3 $R_\mathrm{s}$), the results underestimate the magnetic field at Earth. 
From our results, the calculation of the magnetic field at 1 AU based on NLFFF extrapolations (Fig. \ref{Br_1au}) presents the same underestimation expressed by \cite{2017ApJ...848...70L}. The magnetic field at Earth based on the solution of the NLFFF is slightly closer to the observations compared to the field obtained from the solution of the potential field.
The temporal variation of the magnetic field at 1AU obtained from in-situ measurements (Fig. \ref{Br_1au}a, solid black line) has similar behaviour to the one obtained from extrapolations.
By referring to other studies, \cite{2017ApJ...848...70L} argue that one reason for the underestimation of the interplanetary magnetic field values obtained with the models is the appearance of magnetic switchbacks (the behaviour of the interplanetary magnetic field to fold back on itself). The consequence is the changing of the magnetic field sign that will have the effect of an over-counting field value. \cite{2017ApJ...848...70L} analyzed the open flux for one CR at minimum of activity, and by investigating the sign reversal at Earth, they found a total of 88 hours of this behaviour. Re-evaluating the new observed field they obtained a value with 20\% less than the recorded one. 
In the first part of the cycle (from the beginning of our data till mark number 7 which coincides with the second SSN peak), we see an underestimation of the calculated field at Earth which might be due to the switchbacks. The lowering of the source-surface is not a solution to the discrepancy between the model and the observations. Obviously, by lowering the source-surface to r = 1.8 $R_\mathrm{s}$ we obtain a better estimate than for r = 2.5 $R_\mathrm{s}$. However, this differs from the folding mechanism and a source-surface below 2 $R_\mathrm{s}$ does not agree with field structures concluded from eclipse observations. At the maximum of cycle dynamics (around mark number 8) which differs from the indicated maximum of activity in the SSN we obtain a better agreement between the $|B_\mathrm{x}|$ and $|B_\mathrm{r_{1AU}}|$ compared to the rest of the cycle.


\section{Summary and conclusions}

We applied the NLFFF optimization method on the HMI vector synoptic maps during the solar cycle 24. From the solution of the extrapolation we calculated the unsigned magnetic flux from photosphere to corona for MSOS and sunspot within latitude $\theta=$ [-30$^\circ$, 30$^\circ$], for northern ($\theta = [30^\circ,70^\circ]$) and southern ($\theta = [-70^\circ,30^\circ]$) regions and for northern and southern hemisphere ($\theta = [\pm 70^\circ,0^\circ]$). The free magnetic energy from the NLFFF model indicates the dynamics in the corona, the flare index and flare energy reflects the Sun's dynamics from observations. We calculated the asymmetry indexand the signed flux for the N-S hemispheres. In the corona, the N-S contribution becomes more symmetric at the times with an initial high asymmetry (in the photosphere) and NH dominance. The change in symmetry with height might be due to the variation of the differential rotation with height \citep{2001ApJ...548L..87V, 2010NewA...15..135B}.
From the open magnetic flux maps from the NLFFF solutions at three different heights in the corona (r = 1.4, 1.8, 2.2 $R_\mathrm{s}$), we derive the temporal variation of the magnetic field at 1 AU and compare it with the in-situ measurements. 

We draw a number of conclusions:
\begin{enumerate}
\item The maximum of the cycle activity in terms of emerging flux occurs at the end of 2014, about ten months after the SSN maximum.
\item For the entire solar cycle, $\mathbf{\Phi_\mathrm{MSOS}}$ was the main contributor to the total unsigned flux in the photosphere.
\item The cycle 24 had more sunspots in the NH but a stronger unsigned flux in the SH. 
\item During the decaying phase of the cycle 24, the $\Phi_\mathrm{r}$ from the northern and southern regions presents a sudden drop.
\item After the drop, the flux values remain low compared to $\mathbf{\Phi_\mathrm{r}}$ during the raising and the maximum phase of the cycle.
\item The strongest asymmetries in the unsigned flux occur in all of the atmospheric layers during the Gnevyshev gap (between marks 4 and 5) and after the SSN maximum (mark number 8).
\item The coupling between N and S hemispheres increases with height.
\item The polarity change can vary with height and not only with time.
\item The maximum of the free energy at mark number 8 indicates that during cycle 24, the maximum occurs ten months later than the maximum in the SSN.
\item The Pearson correlation coefficient between the free energy (calculated from the NLFFF model) and the flare index (calculated from observations) is 0.88.
\item The NLFFF method underestimates the magnetic field at 1 AU but slightly less than PFSS models.
\end{enumerate}

\section{Discussions}

The first observation from our data is that the Sun had a different magnetic activity maximum during the solar cycle 24 (marked as number 8 in the Fig. \ref{SSN}, \ref{Flux_hi}, \ref{Efree}, \ref{Flare_index}, \ref{Flares}, \ref{Br_1au}) than the maximum peak of the SSN (mark number 7). We support our conclusion with the results obtained from the temporal evolution of the unsigned magnetic flux which gives us the first indication. The magnetic activity peak (mark number 8) is also  present in the temporal evolution of the free energy, the flare index and in-situ measurements of the magnetic field at Earth. The correlation of 0.88 between free energy and flare index gives us confidence that the NLFFF extrapolation provides us with a quite realistic trend of the solar activity. In the X-ray flare energy data we see that the highest density of the X class flares occurs also in the region marked as number 8. 

Based on differential emission measure (DEM) profiles \cite{2017SciA....3E2056M} created synoptic maps from which they extracted information about temperature (T) and emission measure (EM) in the corona. They compared the T and EM with the radial component of the smoothed photospheric magnetic field. By separating the analysis in the quiet corona and the AR, they conclude that there is neither one-to-one correlation between the photospheric magnetic field and the temperature or with the emission measure. The Fig. 3 (A) of \cite{2017SciA....3E2056M} shows the quiet corona temporal evolution during cycle 24 of $|B_\mathrm{r}|$, T and EM. The temperature has the highest peak right before 2015 near the region marked as number 8 in our plots where the unsigned magnetic flux (Fig. \ref{Flux_hi}), the free energy (Fig. \ref{Efree}) and the flare index (Fig. \ref{Flare_index}) have their highest peak. A rigorous comparison between the solution of the magnetic field extrapolation in the corona and the DEM profiles presented by \cite{2017SciA....3E2056M} might give a better correlation.  

The second conclusion of our results is that the major contribution to the total unsigned flux within the sunspot band latitudes $\theta \in $ [-30$^\circ$, 30$^\circ$] is given by MSOS. The next important contributors to $\Phi_\mathrm{r}$ are the N-S regions (Fig. \ref{Flux_phot_HMI}). The total sunspot flux is one order of magnitude lower than the MSOS flux. This result is in agreement with the study of \cite{2019RAA....19...69J} even though they made a slightly different classification between the flux sources in sunspot/pore and other magnetic structures.
Usually, the dynamo models employ the SSN variation as a reference for the following solar cycle predictions. Including the unsigned magnetic flux activity for these purposes might lead to improved models or predictions.

 From the photosphere to the corona, we notice an oscillatory behaviour with different frequencies and amplitudes in each of the considered categories (Fig. \ref{SSN}, \ref{Flux_phot_HMI}, \ref{Flux_hi}). The flux activity in sunspots and faculae influences the N-S region oscillations in the unsigned magnetic flux. For example, in Fig. \ref{Flux_phot_HMI} at the mark number 7 (activity peak in the SSN), we see one of the peaks in the magnetic flux while in the N-S region variation flux, we notice low values that are probably the consequence of the Gnevyshev gap (mark number 6). To the maximum of the solar cycle dynamics (mark number 8), is contributing the unsigned magnetic MSOS and sunspot flux but it probably also contributes the flux from the N-S regions after it drifted and diffused poleward. On top of this, the super-position of the magnetic shear component over the radial component gives more rise to an enhanced activity right after the SSN peak.
 
  We consider that the reason for the increase in the total unsigned solar and interplanetary magnetic flux visible at mark number 8 is the appearance of the AR 12192 in October 2014. This AR contains "the largest sunspot since November 1990"\footnote{\url{https://www.nasa.gov/content/goddard/tracking-a-gigantic-sunspot-across-the-sun}}. For almost ten days, the AR 12192  produced M and X class flares, but it was poor in CMEs \citep[see for example,][]{2015ApJ...804L..28S}. The effects of its activity were seen at 1 AU for an extended period (Fig. \ref{Br_1au}).

\cite{2017ApJ...848...70L} suggest a couple of reasons for the underestimation of the magnetic flux at 1AU calculated from the extrapolation models. One reason is the observatory maps are systematically underestimating the solar surface magnetic flux. Another argument for underestimation of the interplanetary magnetic field from extrapolations is that, on average, the observed unsigned radial magnetic field is amplified due to switchbacks of the interplanetary magnetic field which can be concluded from frequent sign flips of the radial component. The switchbacks would enhance the unsigned flux without changing the total flux average over large sectors \citep{2017ApJ...848...70L}. These presumably transient field irregularities cannot be modelled by a global field extrapolation. 
If this last mechanism holds, the formation of the folds seems to occur preferentially during the maximum of the SSN. In recent observations by Parker Solar Probe \citep{2019Natur.576..237B}, it was found that the switchbacks of the magnetic fields occur already at 36.6 $R_\mathrm{s}$ at a minimum of solar activity. When a series of strong events as the ones triggered by AR 12192 occur, the fold-backs may be destroyed or inhibited by rapidly propagating transients or by an increase in magnetic pressure. The AR 2192 appeared in the SH. Its dissipation had a strong and immediate effect on the unsigned flux from the SR (Fig. \ref{Flux_phot_HMI}(b), blue curve). In the NR, the dissipation of the unsigned flux was more gentle and occurred at a later time (Fig. \ref{Flux_phot_HMI}(b), red curves.)

\section{Acknowledgements}
Data are courtesy of NASA/SDO and the HMI science teams. This research was supported by the DFG-grant WI 3211/5-1.

\bibliographystyle{aa} 
\bibliography{cmfe}


\end{document}